\documentclass[twocolumn, usenatbib]{mnras}
\pdfoutput=1
\usepackage{graphicx}
\usepackage{natbib}
\usepackage{txfonts}
\usepackage{bm}

\def\apj{ApJ}
\def\apjl{ApJL}
\def\apss{Ap\&SS}
\def\aap{A\&A}

\def\mnras{MNRAS}

\def\beq#1{\begin{equation}\label{#1}}
\def\eeq{\end{equation}}
\def\beqa#1{\begin{eqnarray}\label{#1}}
\def\eeqa{\end{eqnarray}}
\def\eq#1{eq.~(\ref{#1})}

\def\comment#1{\relax}

\title[Comptonization of CMB in galaxy clusters]{Comptonization of CMB in galaxy clusters. Monte Carlo computations}
\author[M. I. Gornostaev and G. V. Lipunova
] {M. I. Gornostaev$^{1,2}$\thanks{E-mail: mgornost@gmail.com} and
G. V. Lipunova$^{1}$\\
$^{1}$ Sternberg Astronomical Institute, Lomonosov Moscow State University, Universitetskij pr., 13, Moscow 119234, Russia\\
$^{2}$ Faculty of Physics, Lomonosov Moscow State University, Leninskie Gory, 1, Moscow 119991, Russia
}

\begin{document}

\date{Received ... Accepted ...}
\pagerange{\pageref{firstpage}--\pageref{lastpage}} \pubyear{2020}

\maketitle

\label{firstpage}

\begin{abstract}
The problem under consideration is to determine the change of the Cosmic Microwave Background  (CMB) spectral shape due to the thermal Sunyaev-Zeldovich effect. We numerically model
the spectral intensity of the  CMB radiation Comptonized by the hot intergalactic Maxwellian plasma. To this aim, a relativistic Monte Carlo code with photon weights is developed.
The code enables us to construct the Comptonized CMB spectrum in a wide energy range.
The results are compared with known analytical solutions and previous numerical simulations. We also calculate the angular distributions of the intensity of radiation emerging from the cloud, which show that the spectral shape of the tSZ effect is not universal for different directions of escaping photons. The numerical method can be applied to simulate the processes of Comptonization for different optical depths, temperatures, initial spectra of photon sources and their spatial distributions, the obtained results may have implications on investigating the profiles of galaxy clusters.

\end{abstract}


\section{Introduction}
\label{s:intro}

The formulation of the Comptonization theory has been extensively developed in astrophysics and cosmology, including studies of galactic X-ray and $\gamma$-ray sources, active galactic nuclei, supernova remnants, and the early Universe.
The first principal progress was made thanks to the introduction of the Kompaneets equation~~\citep{1956Kompaneets}.
This equation describes the evolution of the photon occupation number due to the Compton scattering by thermal electrons  and takes into account  the Doppler shift, recoil, and induced scattering of photons.
An important cosmological application  of the Comptonization theory  has been scrutinized after  discovery of a relative decrease of the CMB temperature in the directions of galaxy clusters~(\citealt{1972AZh....49.1322P, 1973Natur.244...80G, 1973ApJ...184L.105L}; \citealt*{1978Natur.275...40B}). The corresponding problem of modifying the spectrum of a source of low-frequency photons in a hot intergalactic gas has been repeatedly solved analytically and numerically.

The change of the brightness temperature and intensity of the CMB radiation due to scattering by thermal electrons of the intercluster gas was predicted analytically by \cite{1969Natur.223..721S} (see also \citealt{1969Ap&SS...4..301Z})
and is known as the  thermal Sunyaev-Zeldovich effect (tSZ effect).
The main results of this works are derived under the assumption that the parameter $\Theta=k_{\rm B}T_{\rm e}/(mc^2)$ is small, where $T_{\rm e}$ is the electron temperature, $k$ is the Boltzmann constant, $m$ is the electron mass, and $c$ is the speed of light.
Characteristic to tSZ effect is that in the low-frequency range of the spectrum the relative change of the intensity or temperature does not depend on the frequency.
Another effect, the kinematic Sunyaev-Zeldovich effect
\citep{1980PAZh....6..737Z}, is a manifestation of the  cluster motion as a whole and is determined by its peculiar velocity.

It is important to keep in mind other assumptions underlying the Kompaneets equation: the relatively small energy transfer between a photon and an electron in one scattering, the isotropy of the photon space phase function, the non-relativistic Maxwellian distribution of electron velocities,  ignoring of the possible macroscopic motions, inhomogeneity of the medium, and spatial and time variability of the temperature.
Therefore, more sophisticated models (analytical or numerical) of the Comptonization process are necessary for a reasonable comparison  with observations and a correct interpretation of the latter.

For example, the discoveries of a number of  galaxy clusters with high  electron temperatures (more than 15~keV) lead to
an issue of the CMB distortion related to the relativistic effects.
The foremost approaches to this problem are based on the deriving single-scattering profiles of the soft monochromatic photons
(\citealt{1979ApJ...232..348W}; \citealt{1995ApJ...445...33R})
or on the high-order Fokker-Planck considerations of the kinetic equation in order to obtain the relativistic corrections of the Sunyaev-Zeldovich formula (\citealt*{1998ApJ...502....7I}; \citealt{1998ApJ...499....1C}).

The solutions mentioned above describe the behaviour of a model spectrum
in the vicinity of the spectral maximum. Actually,  Comptonization changes the initial spectrum of the soft photons also at the high frequencies,
resulting in a power-law
tail.
The corresponding power-law spectral index in this range was found by \cite{1995ApJ...450..876T}, where the problem about the first eigenfunction of the spatial transfer operator was solved for the spherical and slab plasma clouds with electrons distributed in accordance with the relativistic Maxwellian law. The authors were interested in a factorized solution satisfying the kinetic equation
afar from the characteristic frequency of the photon source.
The slope of spectra at these frequencies was determined as the
function of the Thomson optical depth  of the cloud and the electron temperature.

Observational CMB distortions by hot galaxy clusters with complex geometries and physical conditions can be comprehensively modeled using the Monte Carlo technique. In this paper we present the numerical Monte Carlo simulations which produce the Comptonized spectra in a wide energy range including high-frequency power-law spectra regions.
Here we present the version of the code  that takes into account the thermal Comptonization. We show the consistency  of our numerical results with the existing analytic solutions in the cases where the latter can be applied.  We also calculate for
the angular dependence of the spectral CMB distortions due to thermal Comptonization in a spherical plasma cloud.

In Section \ref{s:analyt}, we present  basic relations describing the Comptonization process and using in the calculations. The physical models, the method and testing of the code are described in Section \ref{s:num}. Results of the computations of CMB modification, including the investigation of angular dependence of spectra and spectral distortion function, are presented in Section ~\ref{s:res} and we provide some concluding remarks in  Section \ref{s:conc}.

\section{Basic relations}
\label{s:analyt}
The thermal Comptonization
is a modification of the radiation spectrum due to inverse Compton scattering of photons by thermal electrons. The basic equations and expressions for quantities in the foundation of the theory of this process have been considered in a number of preceding papers.   Nevertheless, it seems reasonable to summarize them here before presenting our numerical calculations.

\subsection{Scattering}
Let ${\bm \kappa}$ and ${\bm \varv}$ denote the initial photon wavevector and incoming electron velocity
in the laboratory reference frame, respectively, and
$\mathbf{\Omega}={\bm \kappa}/\kappa$ denote the unit vector in the direction of the photon propagation within the solid angle ${\rm d}\mathbf{\Omega}$ before scattering; the primed symbol $\mathbf{\Omega}'$ characterises the direction of the scattered photon.
Let $\eta=\mathbf{\Omega}\cdot\bm{\varv}/\varv$, $\eta'=\mathbf{\Omega}'\cdot\bm{\varv}/\varv$ and $\eta_0=\mathbf{\Omega}\cdot\mathbf{\Omega}'$.
To find the change of the photon frequency after a single scattering one can use the conservation law for the electron and photon four-momenta \citep{1982BLP}.
As a result, the following expression is obtained:
\beq{e:nunu'}
\frac{\nu'}{\nu}=\frac{D}{D'+ \frac{h\nu}{\gamma m c^2}(1-\eta_0)},
\eeq
where $\nu$ and $\nu'$ are the photon frequencies before and after the scattering, respectively,
$D=1-\eta \varv/c$, $D'=1-\eta' \varv/c$, $\gamma$ is the Lorentz factor, $h$ is the Planck constant.  In the case of $\varv=0$, the Doppler shift is zero, and the ratio $\nu'/\nu$ is determined by the recoil effect only.

The differential scattering cross-section is given by the expression
\beq{e:dcrossKN}
\frac{{\rm d}\sigma}{{\rm d}\mathbf{\Omega}'}=\frac{r_{\rm e}^2}{2\gamma^2}\frac{X}{D^2}\left(\frac{\nu'}{\nu}\right)^2,
\eeq
where
\beq{}
X=\frac{\chi}{\chi'}+\frac{\chi'}{\chi}+4\left(\frac{1}{\chi}-\frac{1}{\chi'}\right)+4\left(\frac{1}{\chi}-\frac{1}{\chi'}\right)^2,
\eeq
with
\beq{}
\chi=\frac{2h\nu}{m c^2}\gamma D,~~
\chi'=\frac{2h\nu'}{m c^2}\gamma D',
\eeq
and $r_{\rm e}={e^2}/{(m c^2)}$ is the classical electron radius.

The total scattering cross-section can be expressed by the well-known Klein-Nishina formula,
\beq{e:crossKN}
\sigma(\chi)=\frac{2\pi r_{\rm e}^2}{\chi}
\left\{\left(1-\frac{4}{\chi}-\frac{8}{\chi^2}\right)\ln(1+\chi)+\frac{1}{2}+\frac{8}{\chi}-\frac{1}{2(1+\chi)^2} \right\}.
\eeq

The mean free path is determined by the  scattering probability and for the case of a homogenous medium reads as
\beq{e:lambda}
\langle\lambda\rangle=\frac{\int {\rm d}{\bm p} f({\bm p})}{n_{\rm e} \int  {\rm d} {\bm p} f({\bm p})\,\sigma(\chi) D  },
\eeq
where $n_{\rm e}$ is the electron number density, ${\bm p}$ is the electron momentum, $f({\bm p})$ is the relativistic Maxwellian distribution. Taking into account the isotropy of the latter, the mean relative frequency change due to scattering
can be written as
\beq{}
\frac{{\langle\nu'-\nu\rangle}}{\nu}=\langle\lambda\rangle
\int\left(\frac{\nu'}{\nu}-1\right)\frac{{\rm d}\sigma}{{\rm d}\mathbf{\Omega}'}\, D\, f(p)p^2 {\rm d}p\, {\rm d}\mathbf{\Omega}'\,{\rm d}\mathbf{\Omega}.
\eeq
When both the recoil effect and the Doppler shift are present, one can obtain
\beq{}
\frac{\langle\nu'-\nu\rangle}{\nu}=\frac{4k_{\rm B}T_{\rm e}-h\nu}{mc^2}.
\eeq

\subsection{On some theoretical approaches to the tSZ effect}
The classical approach to the tSZ effect is based on an approximate solution of the Kompaneets equation \citep{1956Kompaneets}.
One considers the regime when
the recoil effect and
induced scattering are of no concern, which takes place if the photon energy $h\nu\ll k_{\rm B}T_{\rm e}$ and the photon
occupation number $n\ll 1$. Then the Kompaneets equation is written as
\beq{e:Komp}
\frac{\partial n}{\partial y}=\frac{1}{x^2}\frac{\partial}{\partial x}\left(x^4\frac{\partial n}{\partial x}\right)\,
\eeq
where $y=\Theta\tau$ is the Comptonization parameter,
with
$\tau$ is the Thomson optical depth of the medium,
and the dimensionless photon energy $x=h\nu/k_{\rm B}T_{\rm r}$ is expressed through the  CMB temperature $T_{\rm r}\simeq 2.7255~{\rm  K}$.

In the case when $y$-parameter is small with respect to unity  one can use the perturbation theory by substituting the undisturbed Planckian value of the occupation number,
$n=({\rm e}^x-1)^{-1}$, in the r.h.s. of \eq{e:Komp}. Then one obtains  the minor distortion of mean intensity $J_\nu=2h\nu^3n/c^2$ \citep{1969Ap&SS...4..301Z},
\beq{e:SZ}
\frac{\Delta J_\nu}{\tau} = I_0\Theta\frac{x^4 {\rm e}^x}{({\rm e}^x-1)^2}\left(x\frac{{\rm e}^x+1}{{\rm e}^x-1}-4\right)\, ,
\eeq
where
$I_0=2(k_{\rm B}T_{\rm r})^3/(hc)^2 \approx 270~{\rm MJy~sr}^{-1}$.
In the Rayleigh-Jeans asymptotic the relative change of the intensity
and brightness temperature of initially blackbody radiation is
\beq{}
\frac{\Delta J_\nu}{J_\nu}=\frac{\Delta T_{\rm RJ}}{T_{\rm RJ}}=-2y.
\eeq
The solution of \eq{e:Komp} for any $y$ comes to
solution of well-known initial value problem for the diffusion equation and
has the form \citep{1969Ap&SS...4..301Z, 1980SvAL....6..213S}
\beq{}
J_\nu(x,y)=\frac{1}{\sqrt{4\pi y}}\int_{0}^\infty J_\nu(x',0) \left(\frac{x}{x'}\right)^3\exp\left(-\frac{(3y+\ln x/x')^2}{4y}\right)\frac{{\rm d}x'}{x'}.
\eeq

The relativistic correction to the solution (\ref{e:SZ}) can be achieved by the Fokker-Planck expansions of higher orders
of the kinetic equation  \citep{1998ApJ...502....7I, 1998ApJ...499....1C, 1998ApJ...508....1S}.
To compare our results with the previous works, we will reproduce the solutions for the intensity distortion of \cite{1998ApJ...502....7I}, which are at the same time in agreement with the solutions presented by \cite{1998ApJ...499....1C}.  The solution including the terms
up to the fifth order in $\Theta$ \citep{1998ApJ...502....7I} has the form
\beq{e:relcorr}
\frac{\Delta J_\nu}{\tau} = I_0\Theta\frac{ x^4 {\rm e}^x}{({\rm e}^x-1)^2}(Y_0+\Theta Y_1 + \Theta^2 Y_2 + \Theta^3 Y_3 + \Theta^4 Y_4),
\eeq
where $Y_0$ is just the expression in the brackets of \eq{e:SZ}  and other $Y$ functions can be found in \cite{1998ApJ...502....7I}.
However, these expansions converge too slowly with increasing temperature due to a widening of the scattering kernel.
The consideration of the effects of multiple scattering
was carried out by \citealt{2001MNRAS.327..567I} in the framework of the Fokker-Planck expansion treatment.
The solutions for arbitrary electron temperature which, additionally,
also take into consideration the multiple scattering,
are obtained by a Monte Carlo modelling of the radiation
transfer \citep{1999ApJ...523...78M} or by a numerical calculation of the Boltzmann collision integral (\citealt{2001MNRAS.327..567I}, \citealt{2001ApJ...554...74D}, \citealt{2012MNRAS.426..510C}).

\section{Numerical computations}
\label{s:num}
\subsection{The models and the method}

To calculate the spectra formed by interaction of low-energy photons with intergalactic hot gas in galaxy clusters we apply a model of spherically symmetric homogeneous medium, possible shocks and local motions of matter are ignored.  Spectral and space distribution of the source photons can be various. Under the Thomson optical thickness of the cloud of radius $R$ we will hereafter
understand  its optical radius $\tau_0 =n_{\rm e} \sigma_{\rm T} R$, where $\sigma_{\rm T}$ is the Thomson scattering cross-section.

Three cases of the photon sources distributions are considered:
to compare with some previous analytical and numerical results, we examine a monochromatic point source at the centre of a homogeneous sphere as well as uniformly distributed sources over the sphere volume;
to model the  CMB Comptionization, we set isotropic blackbody photon sources uniformly on the sphere
surface, and only the photons, with wavevectors initially making the angles less than $\pi/2$ with the direction to the sphere center, are
taken into account.

To model spherically symmetric distributions of particles, it is convenient
to use spherical coordinates and determine the starting points of photon trajectories using
a joint distribution density ${\rm p}(r, \theta, \varphi)$. Since the spherical coordinates of each point are independent, the joint distribution function is factorized. Then each of the quantities $r, \theta, \varphi$ can be modelled independently. The general form of the equations for the calculation of independent random  variable $q$ reads as follows:
\beq{e:Fdist}
F(q)=\xi_i,~~i=1,2,3,
\eeq
where $F(q)$ is a distribution function. Everywhere below, $\xi$
denote the pseudo-random numbers, i.e. independent values of the random quantity distributed uniformly in the interval from 0 to 1. For their generation the C standard library function {\tt rand()} is used. For example, in the case of a uniformly distributed points, equations (\ref{e:Fdist}) lead to the following relations for the modelling of random spherical coordinates $r, \theta, \varphi$
\beq{}
r=\sqrt[3]{\xi_1},~\cos{\theta}=2\xi_2-1,~\varphi=2\pi\xi_3.
\eeq

The logarithmic  energy grid  is determined as $h\nu_j=10^{(N_1+{\rm h}_\nu(j-1))}$ [eV],
where $10^{N_1}$ is the minimum energy, ${\rm h}_\nu$ is the step,  and $j=1...N_\nu$.

The energy of the injected photon is modelled in accordance with the number density distribution, corresponding to the Planckian spectrum,
\beq{}
\mathrm{p}(\nu)=\frac{ h^2\nu^2}{2\zeta(3)(k_{\rm B}T_{\rm r})^3} \frac{1}{{\rm e}^{\frac{h\nu}{k_{\rm B}T_{\rm r}}}-1},
\eeq
where $\zeta(3)$ is the value of the Riemann zeta function.

Let $\mathbf{\Omega}_i$ is the photon direction after the $i$-th scattering.
The random initial photon direction $\mathbf{\Omega}_0$  ($i\!=\!0$ at the initial point of the trajectory)
is chosen in accordance with relations, following from the equations (\ref{e:Fdist}),
\beqa{e:randdir}
\Omega_3=2\xi_1-1,~\Omega_1=\sqrt{1-\Omega_3^2}\cos (2\pi\xi_2),\\\nonumber
\Omega_2=\sqrt{1-\Omega_3^2}\sin (2\pi\xi_2).
\eeqa

In the situation when a small optical depth is considered and a direct Monte Carlo method is inappropriate because of a great number of trajectories required, we use the algorithm based on the computation of the photon weights \citep*{1983ASPRv...2..189P, 1999ApJ...523...78M}.
After energy, position and direction of the photon are set, its mean free
path $\langle\lambda_i\rangle$
 is calculated numerically in accordance with (\ref{e:lambda}).
The distance to the boundary of the cloud  after the scattering which took place at the point with coordinate $\mathbf{r}_i$ is
\beq{}
l_i=-(\mathbf{r}_i\mathbf{\Omega}_i)+\sqrt{R^2-\mathbf{r}_i^2+(\mathbf{r}_i\mathbf{\Omega}_i)^2},
\eeq
and the photon escape probability is
\beq{}
L_i=\exp(-l_i/\langle\lambda_i\rangle).
\eeq
The radius of the cloud $R$ is equal to unity in our computations.
The  weight of the scattered part of the photon is $w_{i+1}=w_i(1-L_i)$, with $w_0=1$; the free path is
$\lambda_i=-\langle\lambda_i\rangle\ln(1-\xi(1-L_i))$, where $\xi$ is a random number, and coordinate of the next scattering is $\mathbf{r}_{i+1}=\mathbf{r}_i+\lambda_i\mathbf{\Omega}_i$.

The simulation of each scattering event begins with modelling of a scattering electron whose direction can be set making use the relations (\ref{e:randdir}).

The value of the electron momentum is modelled in accordance with the relativistic Maxwellian distribution, the density of which in terms of the dimensionless electron temperature $\Theta$ and momentum $z=\frac{p}{m c}$ has the form
\beq{}
\mathrm{p}(z)=\frac{z^2}{\Theta K_2(1/\Theta)}\exp\left(-\frac{\sqrt{1+z^2}}{\Theta}\right),
\eeq
where $K_2(1/\Theta)$ is the modified Bessel function of the second kind of order two, and $0<z<\infty$.
Following \cite{1983ASPRv...2..189P}, to rise the speed of the code performance,  we use
a rejection technique, separately in the cases of low and high temperature of the plasma.
The electron under consideration is accepted as a scattering particle if it satisfies the following selection condition
\beq{}
\xi<\frac{3}{16\pi r_{\rm e}^2}\sigma(\chi)D,
\eeq
where $\xi$ is a random number.

Now it is necessary to obtain the direction of the photon propagation and its energy after the scattering. Let
$\phi'$  be the azimuthal angle being counted in the plane perpendicular to the electron velocity vector (see \cite{1983ASPRv...2..189P} for details
concerning the determination of corresponding reference frame). The joint probability density function of the random variable $\eta'$ and $\phi'$  can be presented in the form
\beq{e:mu-phi}
\mathrm{p}(\eta', \phi')=\frac{1}{\sigma}\frac{{\rm d}\sigma}{{\rm d}\mathbf{\Omega}}=\frac{r_{\rm e}^2}{\sigma}\frac{X}{2\gamma^2 D'^2}\left(\frac{\chi'}{\chi}\right)^2.
\eeq
Then the possible direction of the photon after choosing new $\xi_1$ and $\xi_2$ is
\beq{}
\eta'=\frac{\varv/c+2\xi_1-1}{1+(\varv/c)(2\xi_1-1)},~\phi'=2\pi\xi_2.
\eeq

The algorithm to model $\eta'$, $\phi'$  is dictated by a theorem underlying the von Neumann method of calculation of random quantities,
whose probability density has the form  (\ref{e:mu-phi}), i.e. adds up to the product of
some another probability density and  bounded function.

After calculating the quantity $Y=X^2(\chi'/\chi)^2\leqslant 2$ we choose the direction $\eta', \phi'$ as a new photon direction if $Y>2\xi$. Otherwise
the procedure of the choice of  $\eta', \phi'$ repeats. The new value of the photon energy is calculated in accordance with \eq{e:nunu'}.

Therefore, for each trajectory with number $s$ we obtain the set of quantities
\beq{e:Ijs}
\tilde{I}_j^s=\sum\limits_{i=0}^\infty w_i L_i \delta_{i,\,j},
\eeq
where
\beqa{}
\delta_{i,\,j}=1,&~~\nu_i \in [\nu_{j-1}, \nu_{j}),\\\nonumber
\delta_{i,\,j}=0,&~~{\rm otherwise}.
\eeqa
Then the summation over all trajectories, the number of which is equal to $N_t$, gives the spectrum
\beq{e:sp}
\tilde{I}_j=\frac{1}{N_t}\sum\limits_{s=1}^{N_t} \tilde{I}_j^s,
\eeq
where the quantities $\tilde{I}_j$ are proportional to the intensities since the logarithmic energy grid is used, and their normalization is such that
\beq{}
\sum\limits_{j=1}^{N_\nu} \tilde{I}_j=1.
\eeq
Calculated in such a way quantities $\tilde{I}_j$  can be related with the mean intensity.
In \S\ref{ss:res2} we consider the problem of angular distribution of the Comptonized radiation.

\subsection{The test problems}

For the verification of the code we have solved several benchmark problems and compared the results with earlier analytical and numerical solutions.
\begin{figure}
 	\begin{center}
  \includegraphics[width=0.45\textwidth]{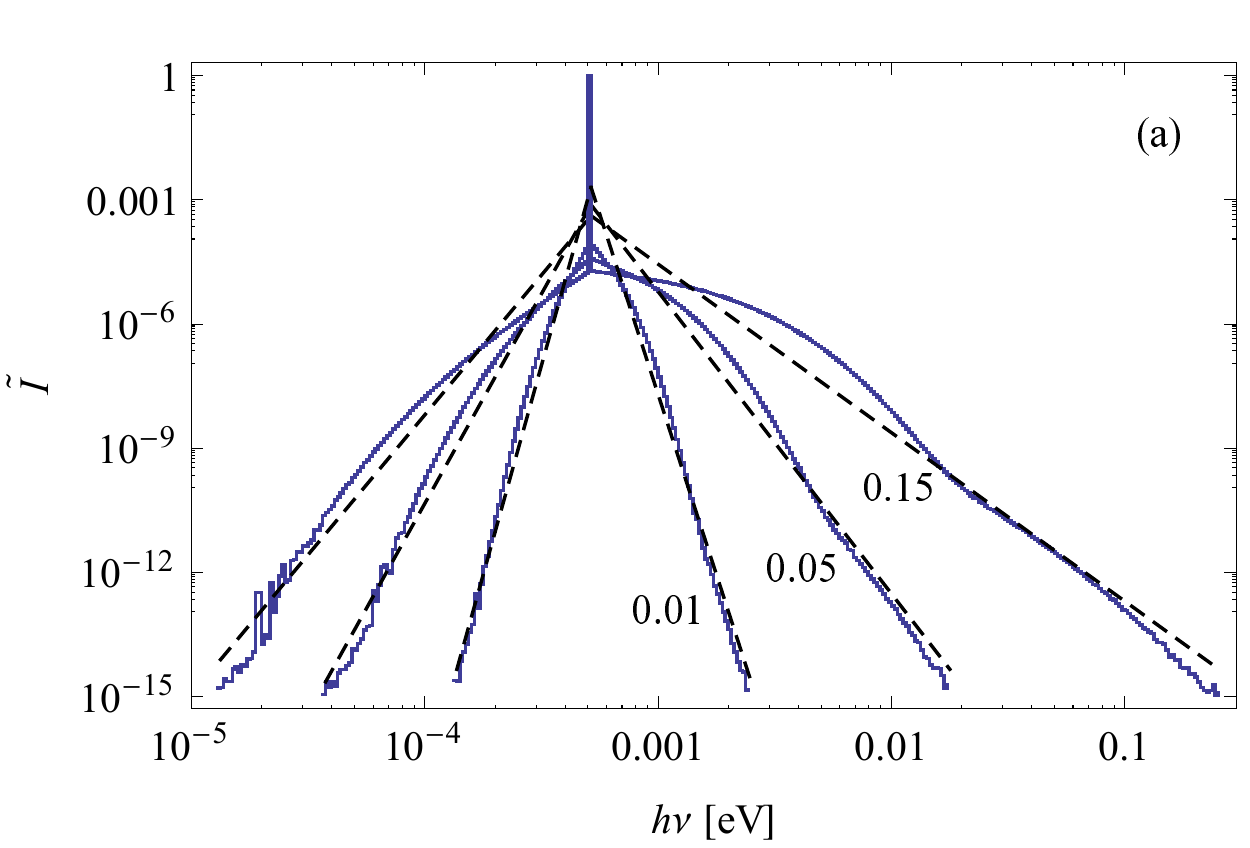}
  \vfill
  \includegraphics[width=0.45\textwidth]{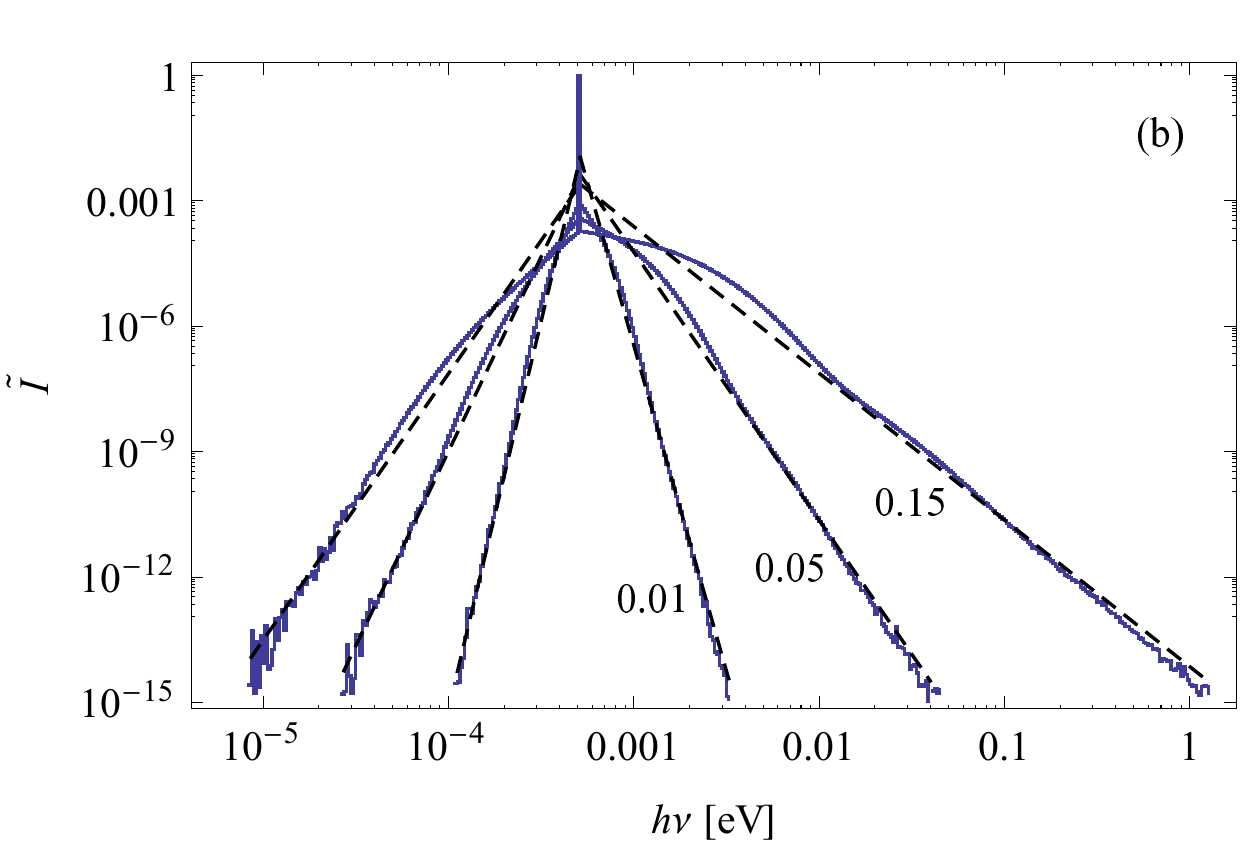}
  \vfill
  \includegraphics[width=0.45\textwidth]{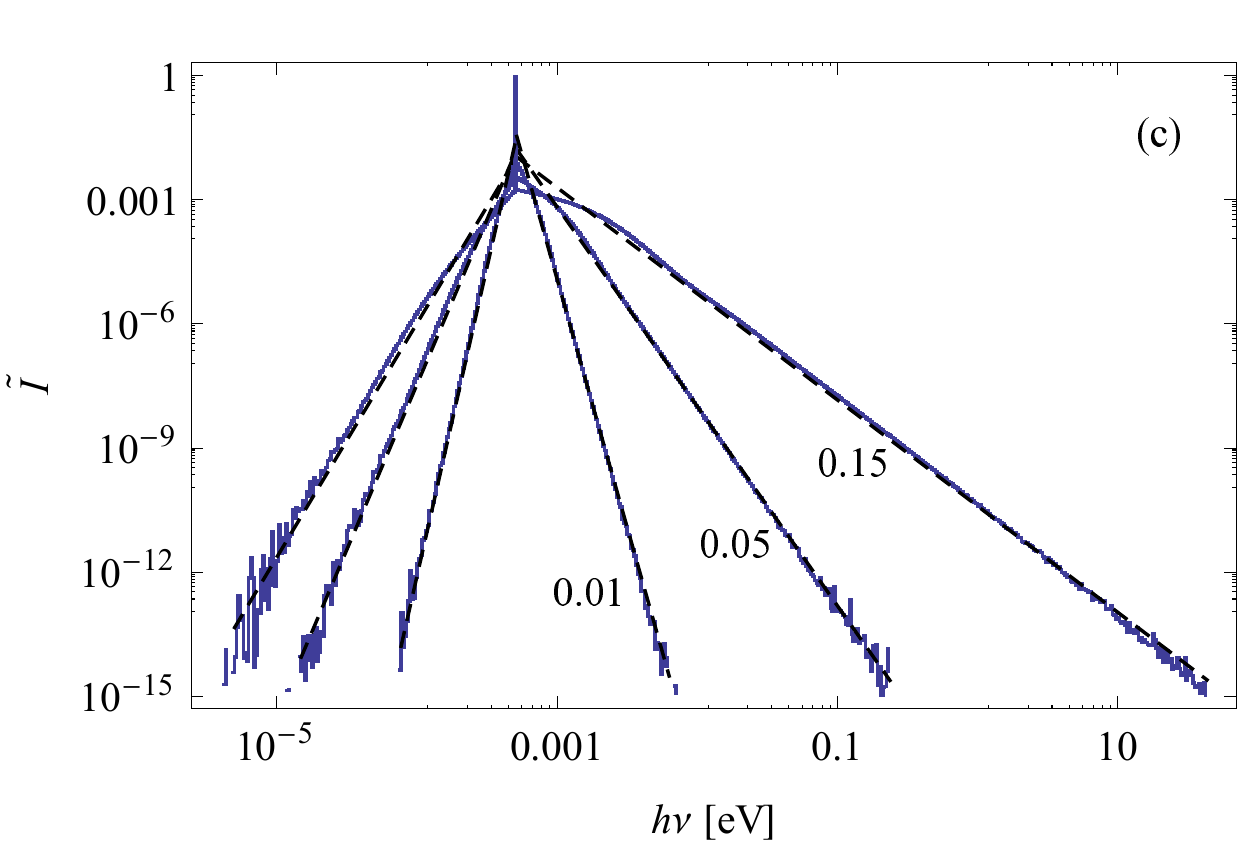}
  		\caption{Comptonization of the monochromatic photons with the initial energy $10^{-9} m c^2$. The photon sources are uniformly distributed over the volume of an optically thin spherical cloud with the Thomson optical depth (a) $\tau_0=10^{-3}$, (b) $\tau_0=0.01$, (c) $\tau_0=0.1$. Monte Carlo results are shown by the blue curves, and their fits  by the broken power laws corresponding to the Green function (\ref{e:GF}) are shown by the dashed lines. The electron temperature in units of $mc^2$ is indicated next to each curve.}
  \label{fig:line}
	\end{center}
 \end{figure}

\subsubsection{Monochromatic low-frequency uniformly distributed photon sources}

If the energy of the soft injected photons is fixed, the behaviour of the solution in this case is determined generally by the electron temperature and optical thickness of the cloud. For the optically thick non-relativistic plasma the problem was  solved analytically by \citet{1980A&A....86..121S}. When the recoil and induced scattering are neglected,  the analytical solution is presented by the convolution of the initial spectrum $\tilde{\nu}_0\delta(\tilde{\nu}-\tilde{\nu}_0)$, where $\tilde{\nu}=h\nu/k_{\rm B}T_{\rm e}$, 
$\tilde{\nu}_0$ is the frequency of the source photons, with the Green function
\beqa{e:GF}
G(\tilde{\nu},\tilde{\nu}_0)=\frac{\alpha(\alpha+\varsigma)}{2\alpha+\varsigma}\frac{1}{\tilde{\nu}_0}\left(\frac{\tilde{\nu}}{\tilde{\nu}_0}\right)^{\alpha+\varsigma},~~\tilde{\nu}\in[0,\tilde{\nu}_0],\\\nonumber
G(\tilde{\nu},\tilde{\nu}_0)=\frac{\alpha(\alpha+\varsigma)}{2\alpha+\varsigma}\frac{1}{\tilde{\nu}_0}\left(\frac{\tilde{\nu}}{\tilde{\nu}_0}\right)^{-\alpha},~~\tilde{\nu}\in[\tilde{\nu}_0,\infty),
\eeqa
where $\varsigma=3$.
Then the mean intensity of the emergent radiation (in dimensionless units) in this particular case is given by
\beq{e:IGF}
J_\nu(\tilde{\nu},\tilde{\nu}_0)=\tilde{\nu}_0\,G(\tilde{\nu},\tilde{\nu}_0),
\eeq
that is fulfilled exactly for the photon sources, distributed in accordance with eq. (8) from \citet{1980A&A....86..121S}.
In the same paper authors notice that for the translucent and optically thin media ($\tau<3$)  the emergent spectra obtained through the convolution diverge from the Monte Carlo spectra obtained by \cite*{1979SvAL....5..149P}, while both approaches are in agreement for the optically thick cases.

In  Fig.~\ref{fig:line}, the Monte Carlo spectra of initially monochromatic photons, obtained in the present work, are showed for different plasma temperatures and optical depth (a) $\tau_0=10^{-3}$, (b) $\tau_0=0.01$, (c) $\tau_0=0.1$.
The dependencies are plotted against the photon energy  together with the results of the fit by a broken
power law $J_\nu=C_1 \ell_1 (\tilde{\nu}/\tilde{\nu}_0)^{\alpha_-} + C_2 \ell_2 (\tilde{\nu}/\tilde{\nu}_0)^{-\alpha_+}$, where $C_1$ and $C_2$ are constants;
$\ell_1=1$, when $\tilde{\nu}\in[0,\tilde{\nu}_0]$, and $\ell_1=0$ otherwise; $\ell_2=1-\ell_1$.
For the case of $C_1=C_2$, corresponding to the fit by the broken power-law Green function,
the obtained values of the parameters $\alpha_-$ and $\alpha_+$
are presented in  Table~\ref{tab:param}. During the fitting we exclude the points lying in the low- and high-frequency ranges, where the relative dimensionless intensity is extremely low (less than $10^{-16}$) and deviates from the power law.
Notice that in the case of $C_1\neq C_2$, when the left and right line wings are fitted separately, we obtain the similar values of the parameters.

\begin{table}
\centering
\caption{Best-fit values of the parameters obtained by fitting the line profiles, presented in Fig.~\ref{fig:line}. }
\label{tab:param}
\begin{tabular}{cccl}
\hline
$\tau_0$ & $\Theta$ & $\alpha^+$ & $\alpha^-$\\
\hline
$10^{-3}$ & 0.01 & 17.32 & 20.29\\
        & 0.05 & 7.32 & 10.11\\  
        & 0.15 & 4.07 & 6.73\\
\hline
$10^{-2}$ & 0.01 & 15.59 & 18.56\\
        & 0.05 & 6.45 & 9.3\\
        & 0.15 & 3.56 & 6.35\\
\hline
$10^{-1}$ & 0.01 & 12.02 & 14.94\\
        & 0.05 & 4.82 & 7.88\\
        & 0.15 & 2.57 & 5.65\\
\hline
\end{tabular}
\end{table}

It can be seen from Fig.~\ref{fig:line} that the role of the multiple scatterings increases with $\tau_0$. For larger $\tau_0$ (Fig.~\ref{fig:line}c) the Monte Carlo spectra conform better  to the analytical solution (\ref{e:IGF}): the problem progressively converges to the diffusion one.

\begin{figure}
\begin{center}
\includegraphics[width=0.45\textwidth]{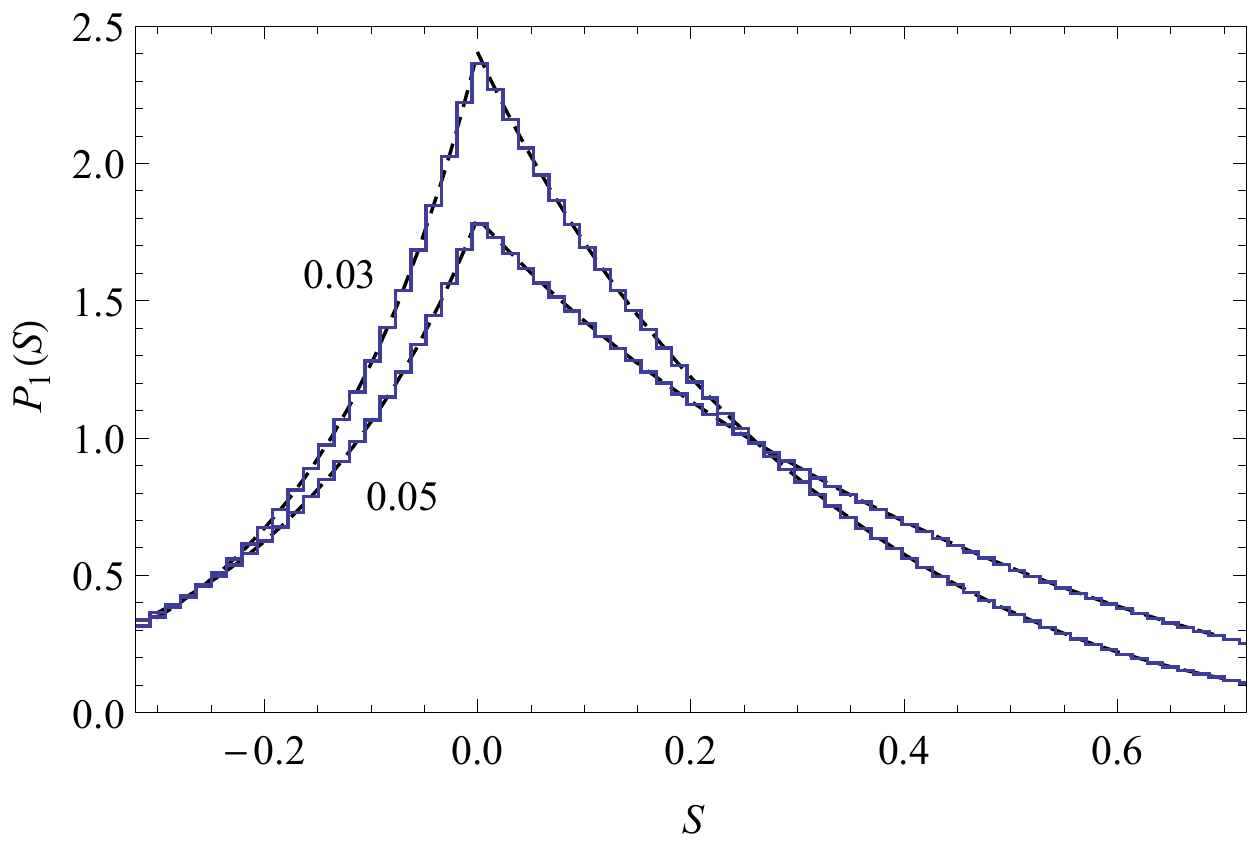}
\vfill
\includegraphics[width=0.45\textwidth]{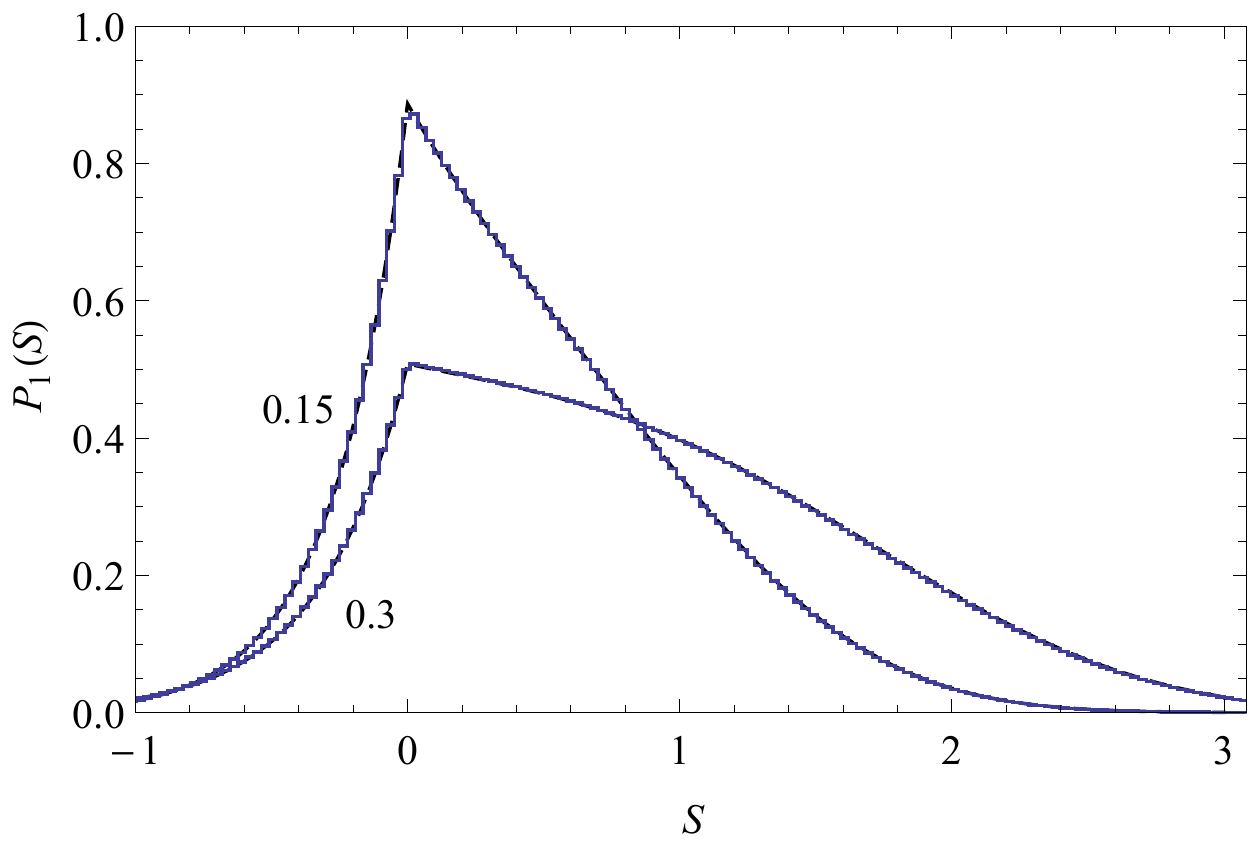}
\caption{The line profile $P_1(S)$ formed by once-scattered monoenergetic soft photons.
Dashed black lines correspond to the analytical expression for $P_1(S)$ (eq. (19) from \citealt{1999ApJ...523...78M}), blue histograms show the Monte Carlo spectra. The electron temperature $\Theta$ is indicated near the curves.}
\label{fig:ssapprox}
\end{center}
\end{figure}

In the optically-thin medium, the contribution from the photons undergoing $N\gg 1$ scatterings is insignificant in the total emergent power.
Actually, they produce the power-law part of the spectrum far from the initial line energy, where the relative intensity is very low (see also \citealt*{1979A&A....75..214P}).
The spectrum near the initial line energy is formed mainly by the photons undergoing a few scattering events  (Fig. \ref{fig:line}a, b).
The narrow peak is formed by unscattered photons.

 The equality
\beq{e:sumtau}
\frac{1}{N_t}\sum\limits_{s=1}^{N_t} \left(\sum\limits_{i=1}^\infty w_i\right)_s = \tau_0
\eeq
is fulfilled in our numerical procedure with a high accuracy for the models of the point source and the surface-distributed sources.
If only the first scattering event is taken into account in the aforementioned sum, the right-hand side changes to $\sim 0.96\tau_0$.

For the  sources uniformly distributed over the sphere volume, the right-hand side of \eq{e:sumtau} equals $\sim 0.77 \tau_0$ when all the scatterings are included, and $\sim 0.71\tau_0$, if only the first scattering is considered.

\subsubsection{Redistribution of photons in a single scattering}

To check the reliability of our numerical computations, we model the function $P_1(S)$, describing redistribution over the frequencies of single-scattered initially monochromatic photons of frequency $\nu_0$ \citep{1995ApJ...445...33R}, with $S=\ln(\nu/\nu_0)$. Details of the  analytical calculation of $P_1(S)$ can be found in \cite{1999ApJ...523...78M}. In  Fig. \ref{fig:ssapprox},  both the numerical and analytical results are presented, being in an excellent agreement.   For computational reasons, the initial energy of the photons is set to $10^{-8} mc^2$. The purpose of this test was to check the algorithms of the present code, which are responsible for the numerical realization of the relativistic Compton scattering kernel.

\subsubsection{Planckian low-frequency central source}

\begin{figure}
\begin{center}
\includegraphics[width=0.45\textwidth]{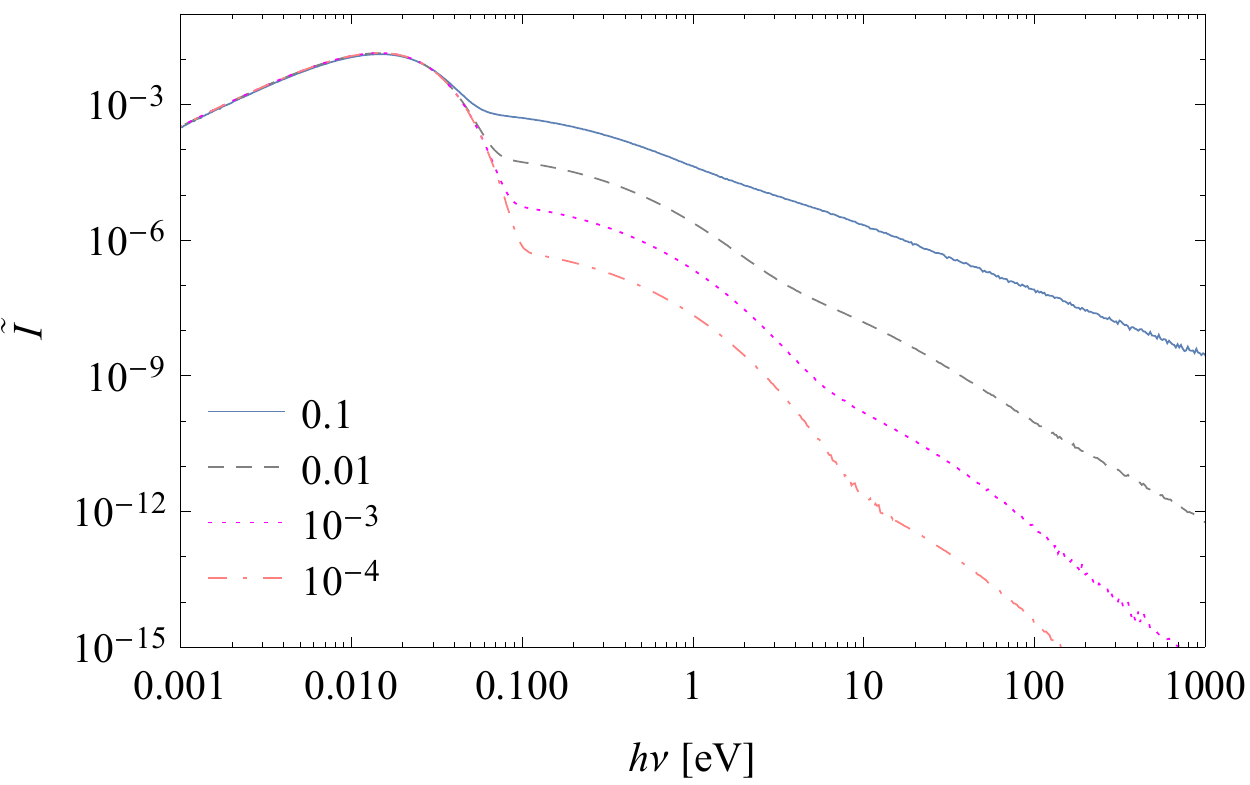}
\caption{Monte  Carlo modelling of the emerged Comptonized spectra of the low-frequency point blackbody photon source with temperature $10^{-8}mc^2$  in the centre of a cloud with relativistic electrons, $\Theta=0.5$. The optical thickness $\tau_0$ is shown in  the legend.}
\label{fig:plank}
\end{center}
\end{figure}

 In  Fig. \ref{fig:plank}, the spectra calculated by the Monte  Carlo code for the spherical cloud with a central point source of photons are presented. The intensity of the seed photons is the Planck function. The results are in  accordance with the earlier calculations of \cite{1983ASPRv...2..189P} (see their figure 37).

\subsubsection{The spectral index: comparison with the solution of the first eigenvalue problem}

The behavior of the power-law spectral index as a function of temperature and optical depth in the problem of Comptonization of soft radiation was investigated by \cite{1995ApJ...450..876T}. The authors were finding a solution of the stationary kinetic equation
\beqa{e:kin}
\mathbf{\Omega}\cdot\nabla I_\nu({\bf r}, \nu,\Omega) =
\int\limits_{4\pi} {\rm d} \Omega' \int\limits_0^\infty {\rm d}\nu'
\left(\frac{\nu}{\nu'}\sigma _s(\nu'\rightarrow\nu,\eta_0,\Theta)I_\nu({\bf r},\nu',\Omega')\right.\\\nonumber
\left.-\sigma_s(\nu\rightarrow\nu',\eta_0,\Theta)I_\nu({\bf r},\nu,\Omega)\right)+S(\nu),
\eeqa
where $I_\nu({\bf r}, \nu,\Omega)$ is radiation intensity,
for the case of $S(\nu)=0$, i.e. for the photon energies far enough from the energy $h\nu_0$ of the source.
Under the assumption of a low-frequency photon source ($h\nu \ll k_{\rm B}T_{\rm e}$ for all injected photons) only the Doppler term can be left,
and the Klein-Nishina scattering kernel has the form
\beqa{e:kernelD}
\sigma_s(\nu\rightarrow\nu', \eta_0,\Theta)=
\frac{3\sigma_{\rm T}}{16\pi}\frac{n_{\rm e}}{\nu \hat{\nu}} \int {\rm d}{\bm \varv} \frac{f(\varv)}{\gamma} \delta\left(\gamma\left(\frac{ D}{\hat{\nu}'}-\frac{ D'}{\hat{\nu}}\right)\right)\times\\\nonumber
\times\left(1+\left(1-\frac{1-\eta_0}{\gamma^2 DD'}\right)^2\right),
\eeqa
where $\hat{\nu}=h\nu/mc^2$.
Here  the dependence on the  initial and final photon directions  is changed for the dependence on the $\eta_0$.

With such  scattering kernel $\sigma_s$ and assuming a power-law spectral distribution, problem (\ref{e:kin}) is reduced to the eigenvalue problem for the spatial operator, which in the optically thick plasma comes to the diffusion one. In the optically thick case, the first eigenfunction, which is broken power-law, is enough to describe the spectral shape  practically at any frequency, even very near the source energy. Furthermore, if the plasma is non-relativistic,  the solution of the problem reduces to the one described in \cite{1980A&A....86..121S}.  In the optically thin case, the first eigenfunction is not sufficient to represent the solution accurately near the initial photon energy, as can be seen, for example, from Fig.~\ref{fig:line}a, b.

\begin{figure}
  	\begin{center}
  \includegraphics[width=0.45\textwidth]{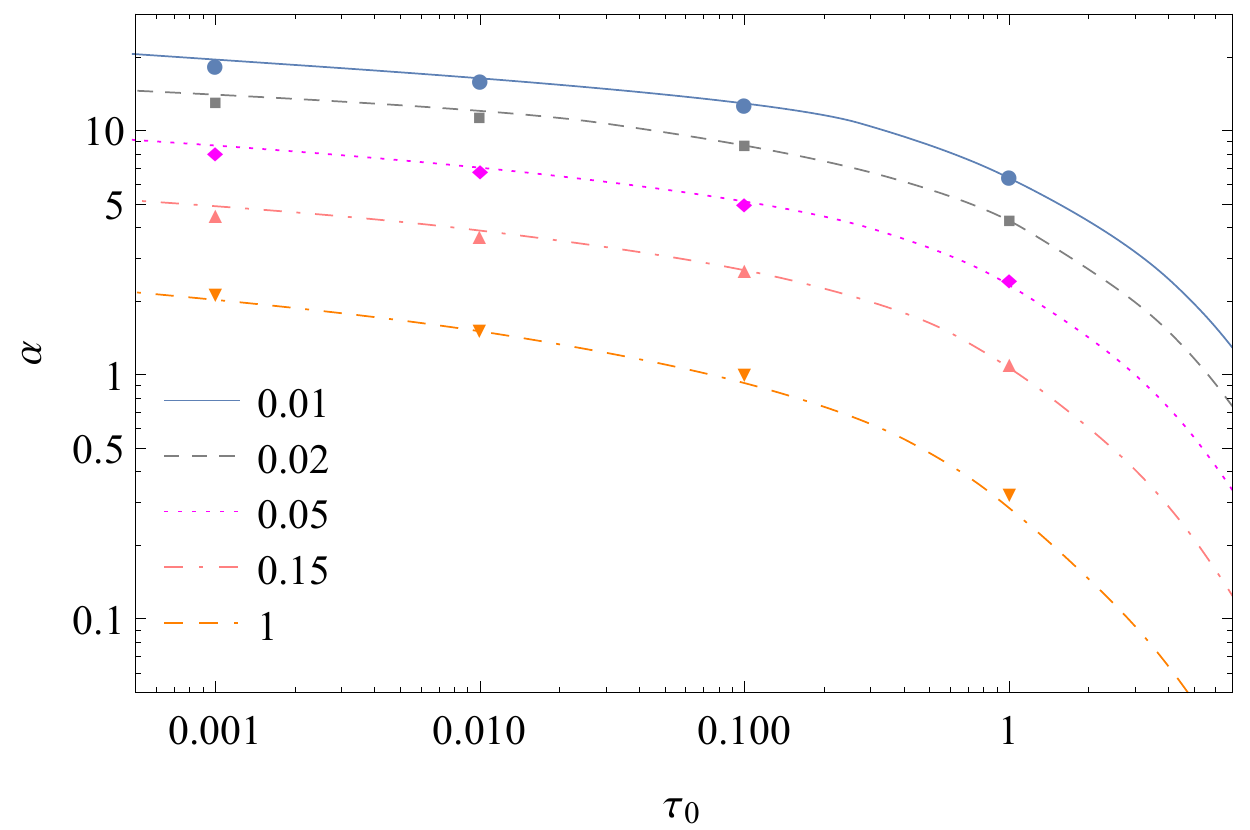}
  		\caption{The power-law spectral index plotted as a function of the optical thickness $\tau_0$. The lines show the analytical values calculated using \eq{e:disp}. The temperature $\Theta$ is indicated in the legend. Each dot corresponds to the fit of the Monte Carlo spectra produced by Comptonization of monoenergetic photons from uniformly distributed sources.}
  \label{fig:dist}
	\end{center}
 \end{figure}

The results of calculations of spectral index $\alpha$ as a function of the optical thickness $\tau_0$ are presented in Fig.~\ref{fig:dist}. The index is found from the dispersion relation  \citep{1995ApJ...450..876T}
\beq{e:disp}
C_0(\Theta, \alpha)\Lambda(\tau_0)=1,
\eeq
where $C_0$ is the first (non-zero) coefficient in the expansion of the phase function
in a series of Legendre polynomials and $\Lambda$ is the first eigenvalue of the radiation transfer integral operator
(see equation (16) of  \citealt{1995ApJ...450..876T}).

Numerical results for $\alpha$ are shown by points in Fig.~\ref{fig:dist}.
Each point represents a fit to a Monte Carlo spectrum produced as a result of Comptonization of the monoenergetic photons in the case of the uniformly distributed sources. As one can see, the spectral indexes of the numerically obtained spectra are in a good agreement with the analytical ones.

Although the intensity in the power-law region of the spectra  is relatively insignificant for $\tau\ll 1$, our results demonstrate the contribution of multiple scattering events in the situation of an optically thin medium and their increasing role
in the spectrum  with growing optical thickness.

 \subsection{Internal parameters of the code}

 Besides the physical parameters of the model, like the optical depth, plasma temperature, and temperature of the injected radiation, there are parameters of the numerical procedure: the cutoff energy, the step of the energy grid, and the number of the trajectories. We choose these quantities depending on the specific situation.

 Since the form of the Comptonized spectrum of a monoenergetic line near the maximum has a complicated shape, it is reasonable to
check the influence of the parameters of the numerical procedure  for all problems under consideration.

The tests
show that the spectrum shape (including the distortion curves, see below) becomes independent of the number of the calculated trajectories $N_t$ starting from some number $N'_t<N_t$ (and thus $N_t$ can be used to  obtain the final results). Moreover, it was checked that the cutoff energy is chosen low enough not to have an effect in the spectral region under examination. A variation of the step of the energy
grid influences the smoothness of the function, obtained by interpolating the histogram (\ref{e:sp}).
The step is chosen in a reasonable manner for a specific problem, taking into account the  required precision and computational possibilities.
Therefore, we believe that the physical adequacy of our numerical results  is quite justified.

\section{Modifications of CMB spectrum}
\label{s:res}
\subsection{Angle-averaged tSZ effect}
\label{ss:res1}

The normalized distortion of the CMB calculated in accordance with  \eq{e:SZ} and \eq{e:relcorr}, does not depend on the cloud optical depth.
Our computations confirm the validity of this statement in the frequency range usually involved in the analysis of the SZ effect and show that it holds independently from the temperature and  the distribution of the incident photons   (see below).

In Fig. \ref{fig:SZmc}, the results of the numerical computations, as well as the solutions (\ref{e:SZ}) and (\ref{e:relcorr}) are presented (hereafter in the figures and the text the intensity and its distortions are
measured in units of $I_0$).
The dots correspond to the quantities $\tilde{I_j}$, given by \eq{e:sp} and taken at the geometrical mean of the energetic boundaries of the bins, i.e., we assume that in these points the histogram values are proportional to the intensity,
$\tilde{I}_{j-1/2}=\tilde{I}_j(\sqrt{\nu_{j-1} \nu_{j}})\propto J_\nu(\sqrt{\nu_{j-1} \nu_{j}})$.
 Hereafter $J_\nu$ is the intensity averaged over the hemisphere outer with respect to the cloud boundary, $J_\nu=\int_0^1 I_\nu(\mu){\rm d}\mu$,
where $I_\nu(\mu)$ is the radiation spectral intensity at the boundary, with
$\mu$ is the cosine of the angle between a given direction
and the outer normal to the sphere at the same point. (Due to the spherical symmetry of the problem,
at each point on the surface the dependence of the intensity on the angle is the same, so
we are not interested in the coordinates of the emitted photons.
The dependence on the azimuthal angle measured in the plane tangent to the sphere at the escape point is also absent, of course.)
The quantity $J_\nu$ characterises only the radiation
outgoing from the cloud (`Comptonized' photons).
Considering the mean intensity at the boundary one can obtain
$J_\nu'=\frac{1}{2}\int_{-1}^1 I_\nu(\mu){\rm d}\mu=(B_\nu+J_\nu)/2$.

Our numerical solutions for the distortion
$\Delta J_\nu=J_\nu-B_\nu=2(J_\nu'-B_\nu)$, being normalized to $\tau_0$,
are in a good agreement with the solution (\ref{e:relcorr}) in the particular temperature range.
Thus, numerical calculations show that the transition  to the deviation of mean intensity $J_\nu'-B_\nu$ at the boundary
can be carried out by multiplying the normalized distortion
$\Delta J_\nu/\tau_0$ by the factor $\tau_0/2$.
The expression \eq{e:relcorr} becomes
inaccurate already at the temperatures about $0.04 mc^2$ (however, the same solution is more accurate
for such temperatures when it includes only
the first four terms, see \citealt{1998ApJ...502....7I}).

In Fig. \ref{fig:SZmc2}a, the solutions in the wider range of temperatures, from $\Theta=0.01$ to $\Theta=0.1$ are plotted for two values of the  optical thickness of the cloud, $\tau_0=0.1$ and $\tau_0=0.05$. Apparently, there is no significant dependence on $\tau_0$.
Since the solutions of the type (\ref{e:SZ}) and (\ref{e:relcorr}) are obtained
under the condition that the $y$-parameter is fixed for all the photons independently from their initial direction,
we compare them also with calculations in the model of the central point source for $\tau_0=0.1$ and $\tau_0=0.05$ (Fig.~\ref{fig:SZmc2}b).
The linearity of the effect on $\tau_0$ seems to be a property independent from the plasma temperature
and from the model of the initial spatial distribution of the photons in the considered range of $\tau_0$.
The minor deviations from linearity on $\tau_0$, registered in the current calculations,  are browsed in Fig.~\ref{fig:SZmc2}.

\begin{figure}
\begin{center}
\includegraphics[width=0.45\textwidth]{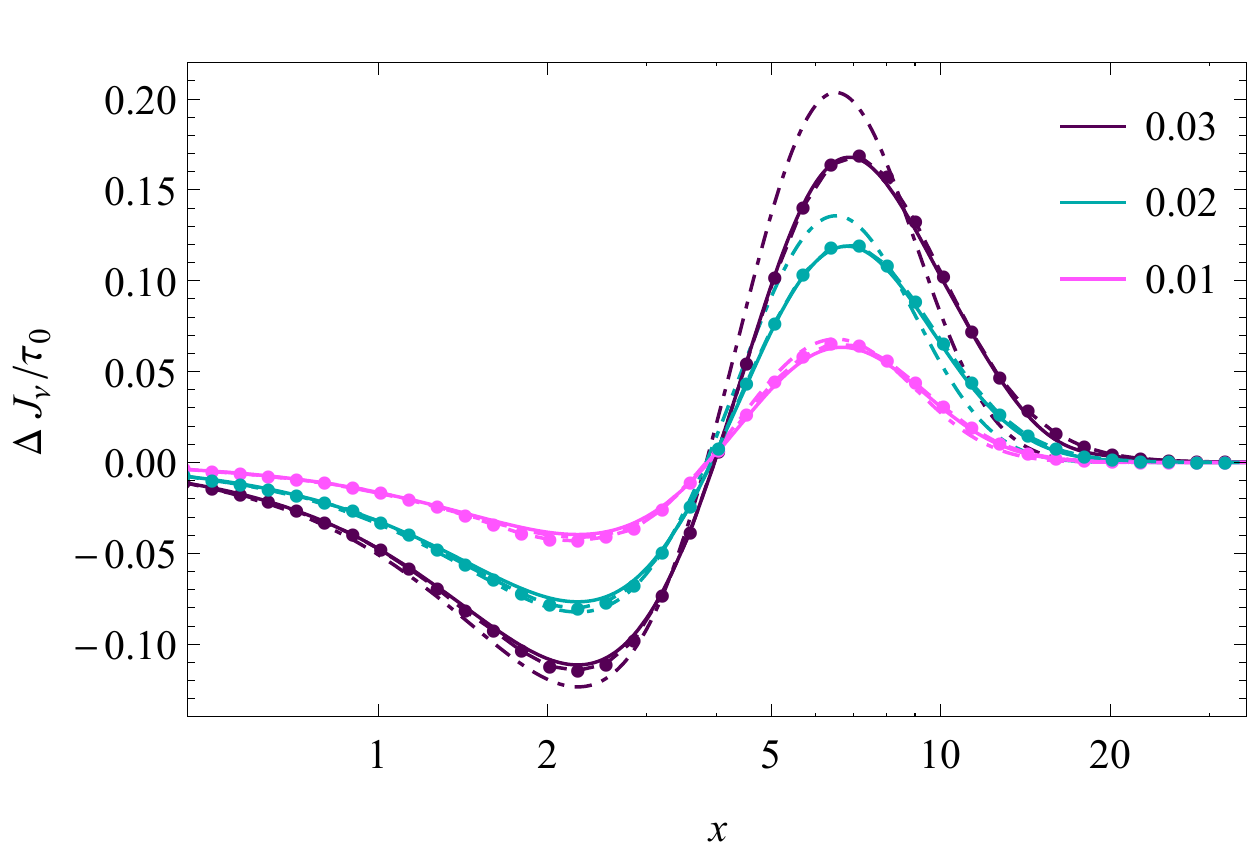}
\caption{The distortion of spectral intensity of the Comptonized blackbody photons, calculated in accordance with \eq{e:SZ} (dot-dashed lines) and \eq{e:relcorr} (solid lines) and as a result of the Monte Carlo computations in the model of the photons sources uniformly distributed on the spherical surface (dots with dashed lines). The dimensionless temperature $\Theta$ is indicated in the legend.}
\label{fig:SZmc}
\end{center}
\end{figure}

\begin{figure}
\begin{center}
\includegraphics[width=0.45\textwidth]{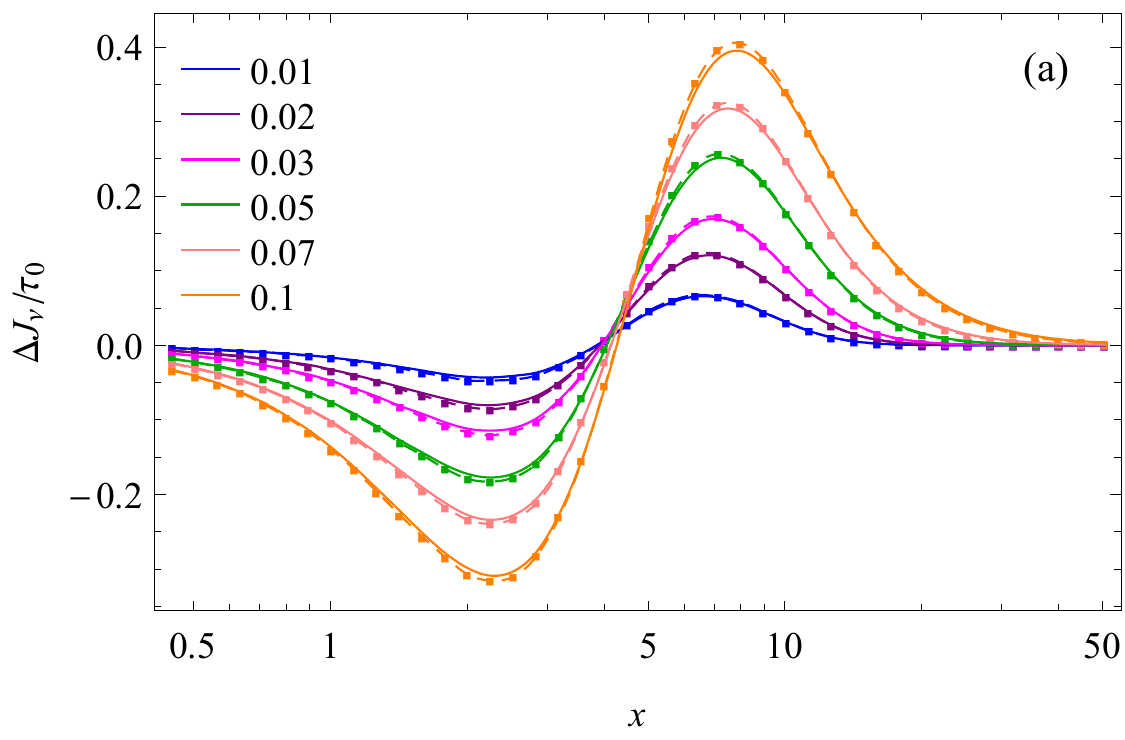}

\includegraphics[width=0.45\textwidth]{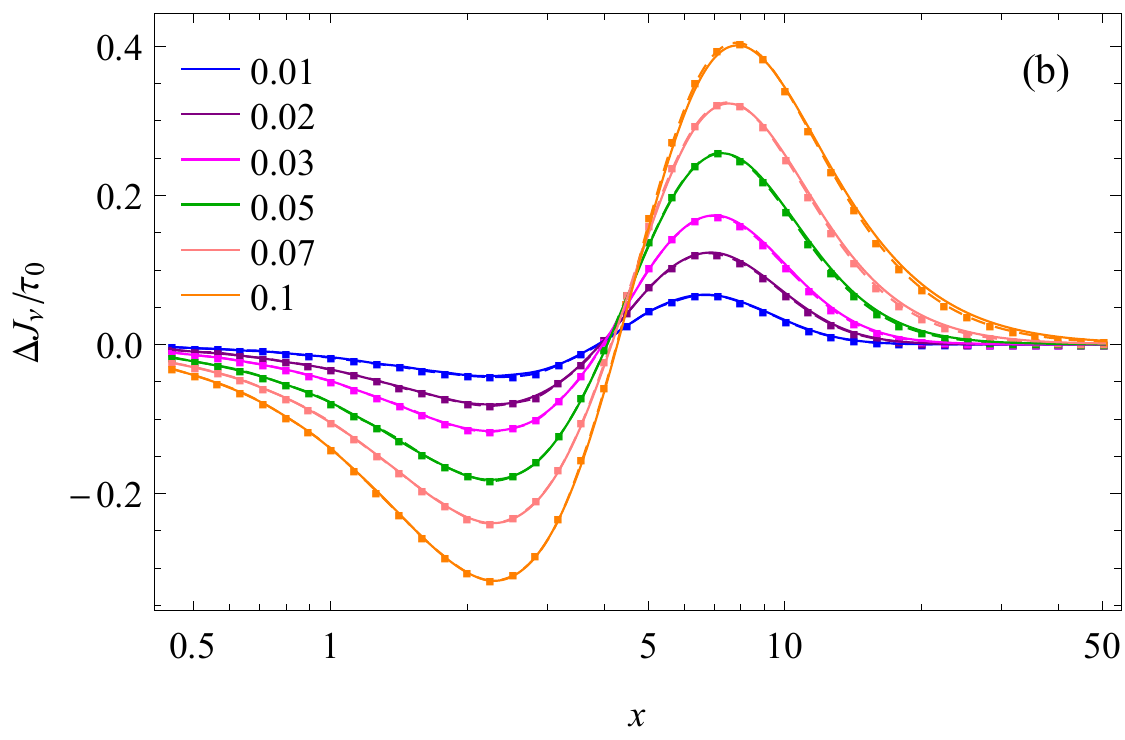}
\caption{The distortion of spectral intensity  of the Comptonized blackbody photons,
calculated numerically for different $\Theta$ from 0.01 to 0.1, indicated for the curves in the legends
(the absolute value of distortion grows with $\Theta$).
The calculations are carried out in the frame of
the boundary-distributed photon sources model (a)  and
the central source model (b),  for optical thickness $\tau_0$ equal to 0.1 (solid lines with circles) and 0.05 (dashed lines).
}
\label{fig:SZmc2}
\end{center}
\end{figure}

\begin{figure}
\includegraphics[width=0.45\textwidth]{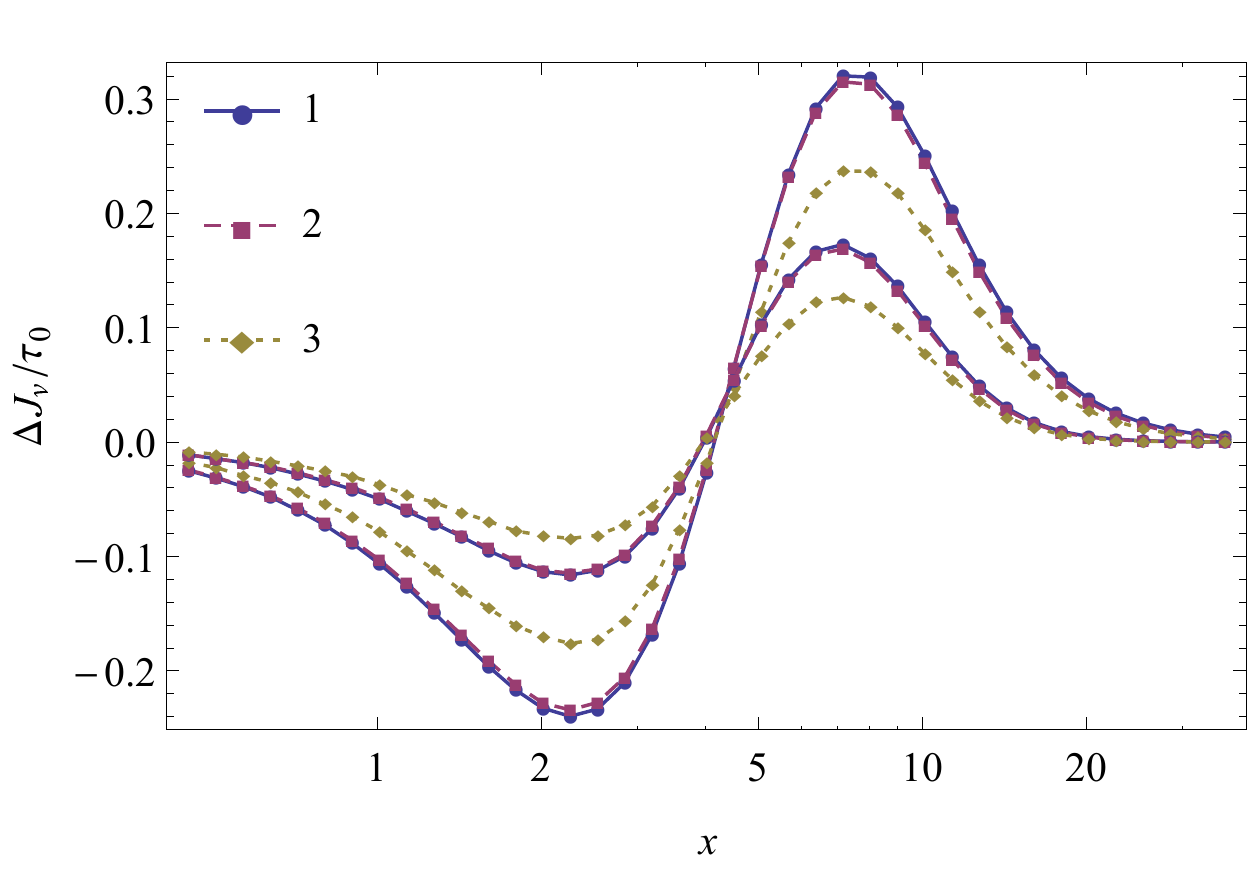}
\caption{Spectral distortions of the CMB, obtained for three different distributions of the blackbody source photons: central source (1), uniformly distributed sources on the sphere surface (2), uniformly distributed sources over the sphere volume (3).  In each case, the curve of lesser amplitude corresponds to $\Theta=0.03$, and the other, to $\Theta=0.07$. The optical thickness $\tau_0=0.1$.
}
\label{fig:3models}
\end{figure}

 For methodological aims,  we have also solved the Comptonization problem for the photon sources uniformly distributed  over the volume of the spherical cloud.
In Fig. \ref{fig:3models}, the results obtained in the three different models of initial spatial photon distributions are compared.
 The amplitude of the
effect for uniformly distributed sources is lessened. The linearity
on $\tau_0$ still takes place near the source frequency.
The ratio of the quantity $\Delta J_\nu$ for the uniform spatial initial photon distribution to the one
in the case of boundary-distributed (or central) source is $\sim  0.74-0.76$ in the vicinity
of the maximal deviation. This approximately corresponds to the coefficient 0.77 in the r.h.s of \eq{e:sumtau}.

\begin{figure}
\includegraphics[width=0.44\textwidth]{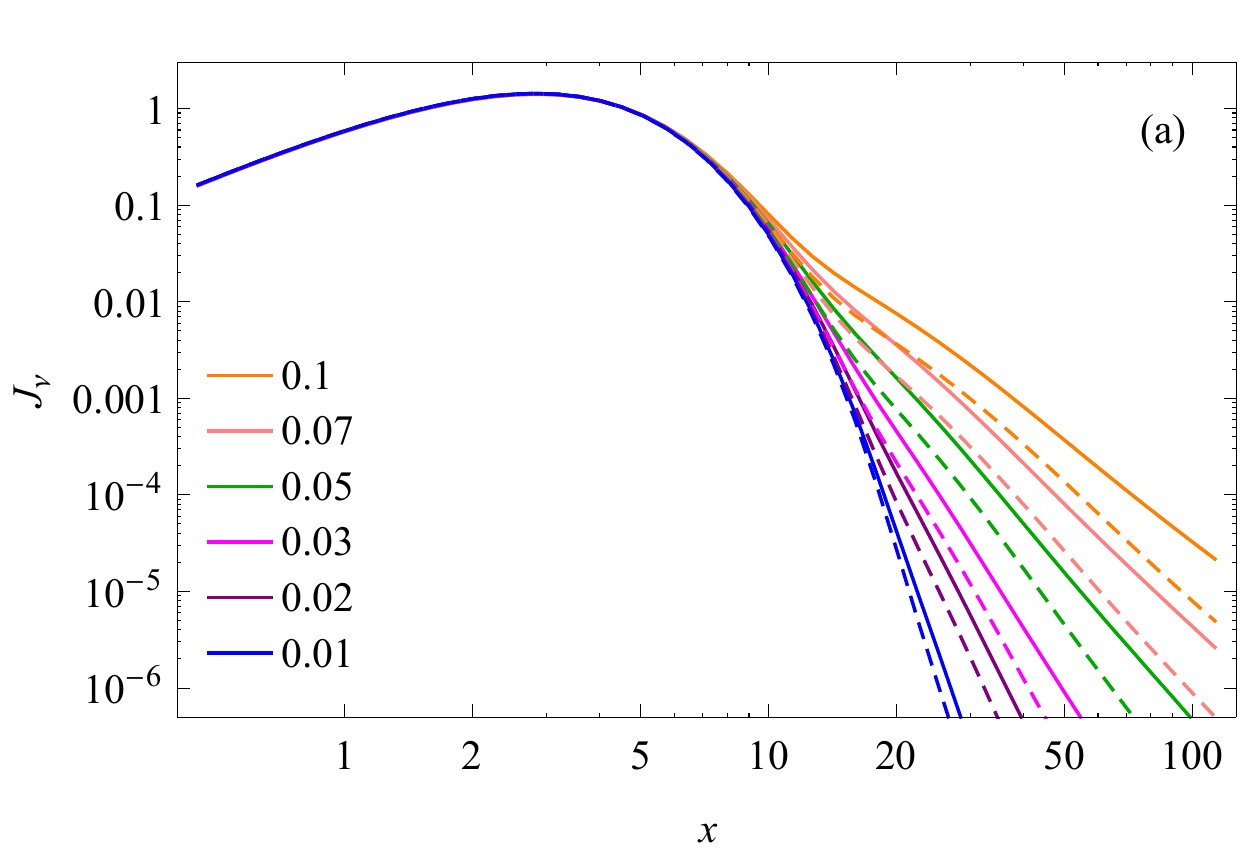}
\includegraphics[width=0.44\textwidth]{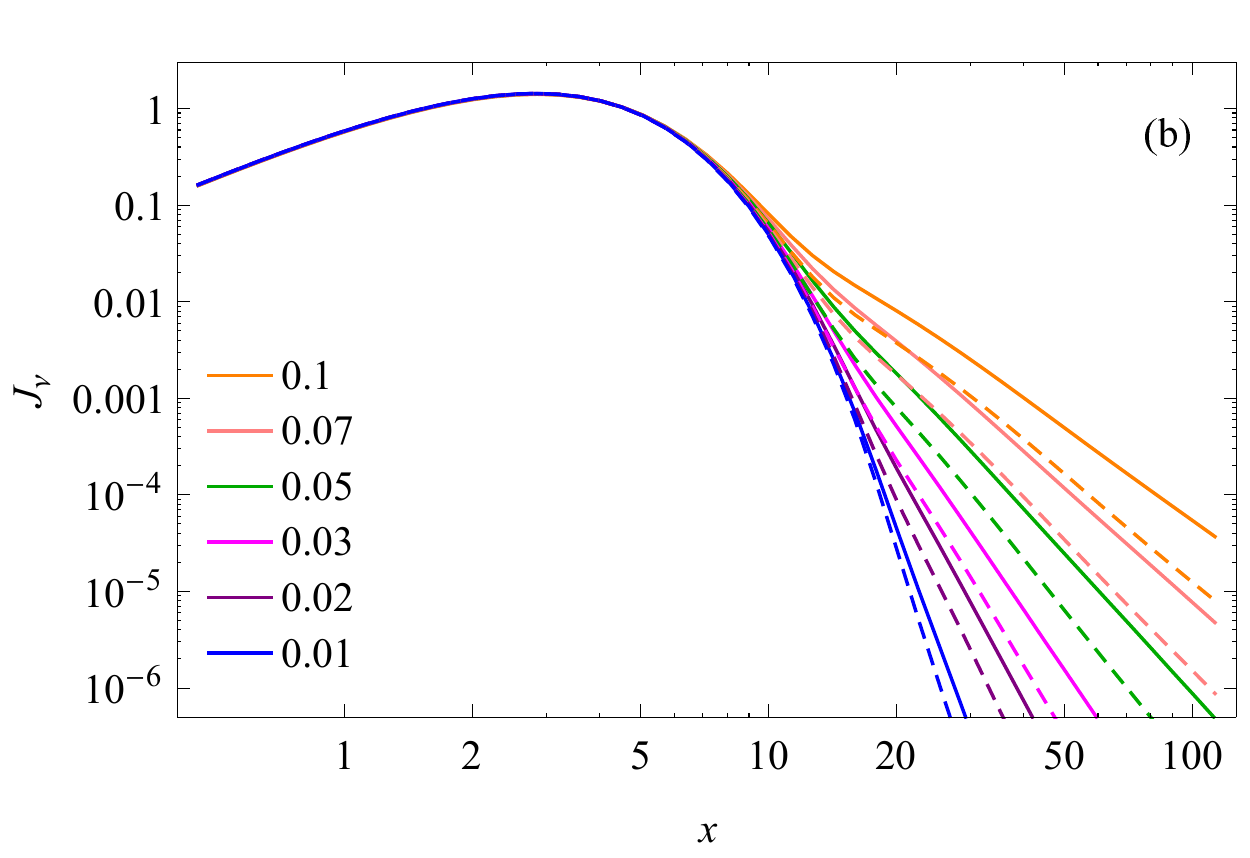}
\caption{Spectra of the CMB after the thermal Comptonization in a galaxy cluster, calculated
 for the optical thickness $\tau_0$ equal to 0.1 (solid lines) and 0.05 (dashed lines) in the boundary-distributed photon sources model (a),
and  in the model of  central source  (b).
The temperature $\Theta$ is indicated in the legends.
}
\label{fig:spec}
\end{figure}

Examples of the calculated spectra, demonstrating the power-law character at high frequencies,
are presented in the Figs. \ref{fig:spec}a and  \ref{fig:spec}b, for the sources at the surface of the sphere and
the central source, respectively (the dot markers
are hereafter dropped from the figures).
The spectrum slope is practically independent from the model (it is also the same for the volume-distributed sources). The values of the spectral index $\alpha^*$, calculated in accordance with \eq{e:disp}, as well as values $\alpha$ obtained by fitting  the power-law part of numerically calculated spectra, can be found in Table~\ref{tab:indexes}.

\begin{table}
\centering
\caption{Best-fit spectral index $\alpha$ obtained by fitting the CMB power-law spectral interval. Model 1 is the photon sources distributed over the spherical surface, and Model 2 is the central point source. The values of $\alpha^*(\tau_0,\Theta)$ are given by the solution of \eq{e:disp}.}
\label{tab:indexes}
\begin{tabular}{ccccc}
\hline
& &\multicolumn{1}{|c|}{Model 1} & \multicolumn{1}{|c|}{Model 2}&\\
\hline
 $\tau_0$ & $\Theta$ &  $\alpha$   & $\alpha$ & $\alpha^*$ \\
\hline
   0.1  &0.01 &  12.73     & 12.39 & 12.68\\
        & 0.02      &8.69      & 8.45 & 8.64 \\
        & 0.03      &6.75      & 6.58 & 6.86 \\
        & 0.05      &5.09      &  5.04 & 5.11 \\
        & 0.07      &4.24      &  4.15 & 4.20 \\
        & 0.1       &3.44      &  3.39 & 3.40 \\
\hline
   0.05  & 0.01  & 14.42     & 14.77 & 14.05 \\
         &0.02      & 9.77    & 9.71 &9.62 \\
         &0.03      & 7.81    & 7.55 & 7.67\\
         & 0.05     & 5.84    & 5.71 & 5.75\\
          & 0.07    & 4.72    & 4.70 & 4.74\\
          & 0.1     & 3.90    & 3.84 &3.86 \\
\hline
\end{tabular}
\end{table}

\subsection{Angle-resolved tSZ effect}
\label{ss:res2}

The calculation of the spectra described above implies the summation of the values (\ref{e:Ijs}) corresponding to all the 'parts' of photons emitted from the cloud, regardless of the escape directions.
We now ask ourselves about the angular distribution of the spectral intensity of radiation transmitted through the cloud, $I_\nu(\mu)$.
Let us  consider in this subsection solely the model of the surface-distributed sources and introduce an additional uniform grid for the value of $\mu$: $\mu_k=k{\rm h}_\mu$, $k=0...10$, ${\rm h}_\mu=0.1$.

\begin{figure*}
\begin{center}
\includegraphics[width=0.45\textwidth]{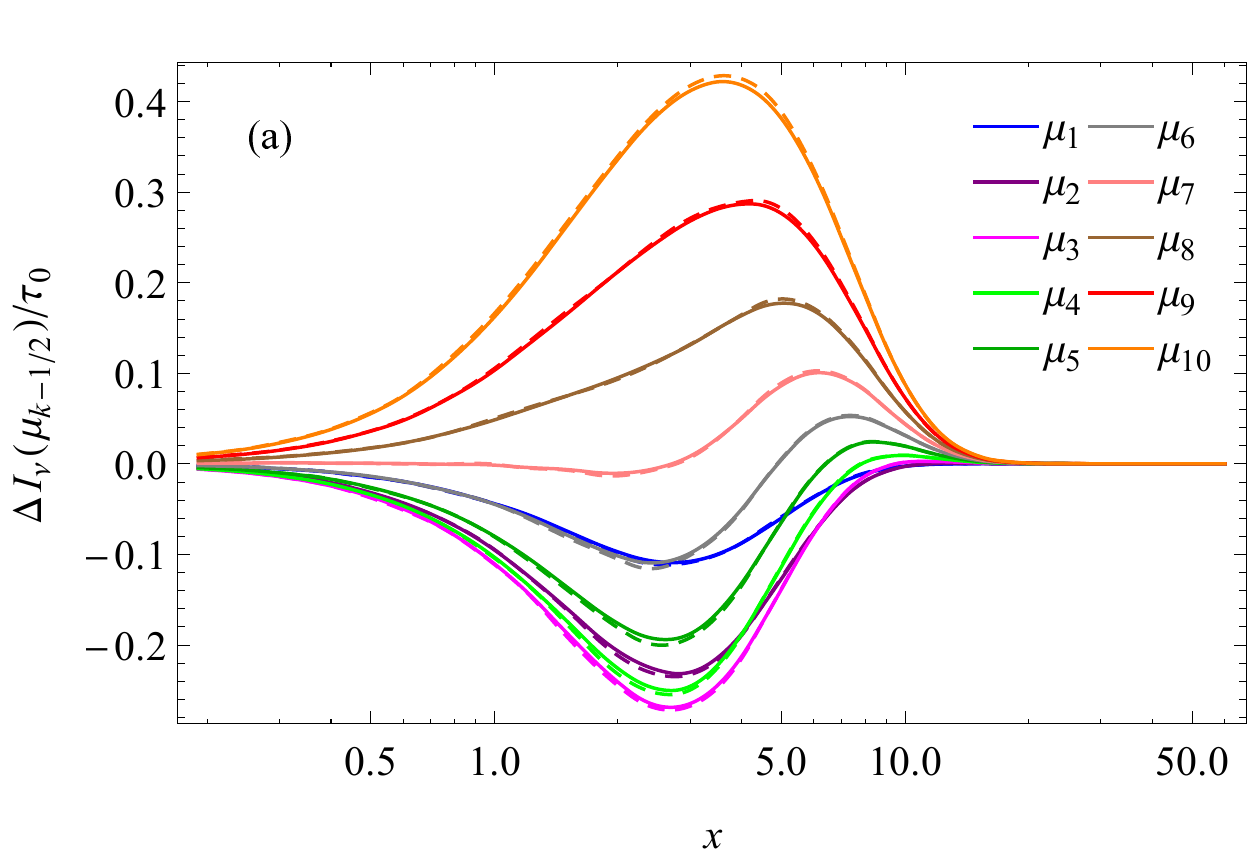}
\hfill
\includegraphics[width=0.45\textwidth]{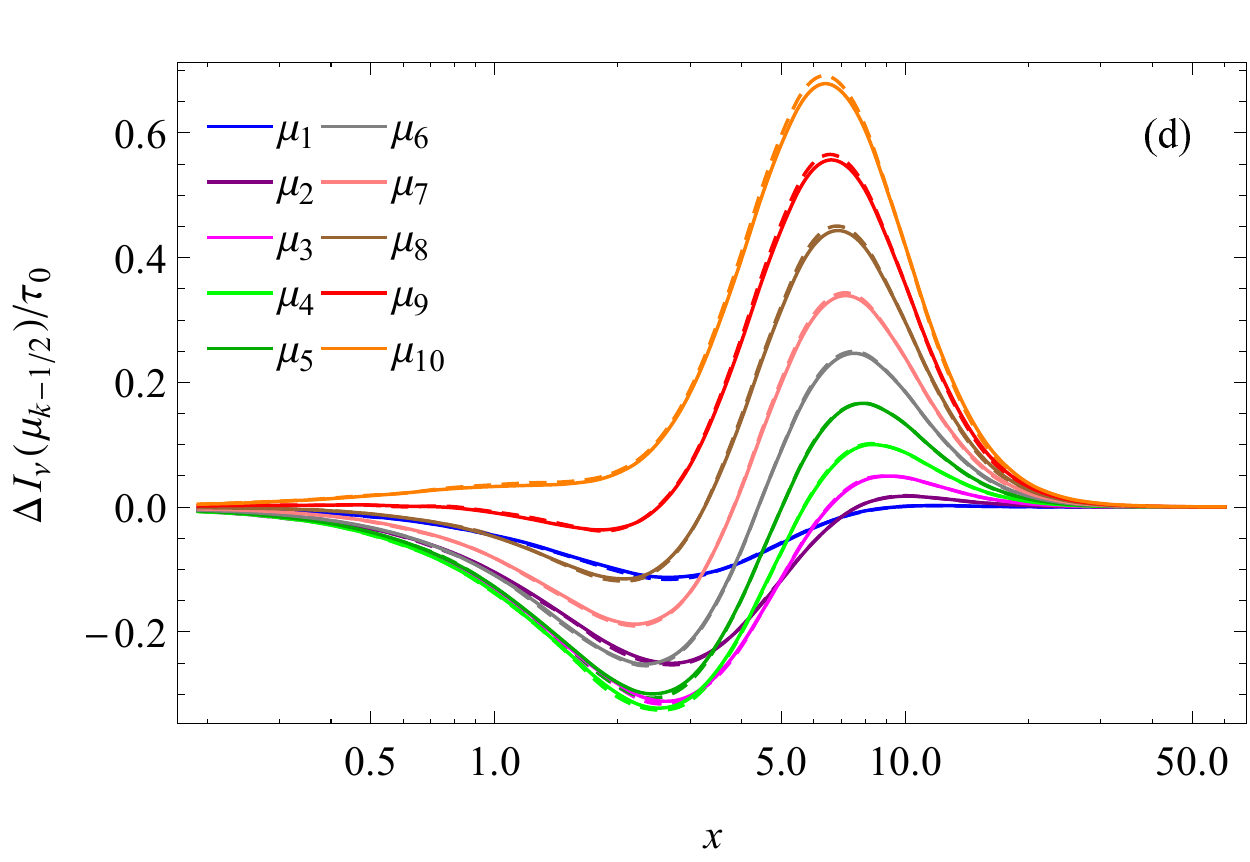}
\vfill
\includegraphics[width=0.45\textwidth]{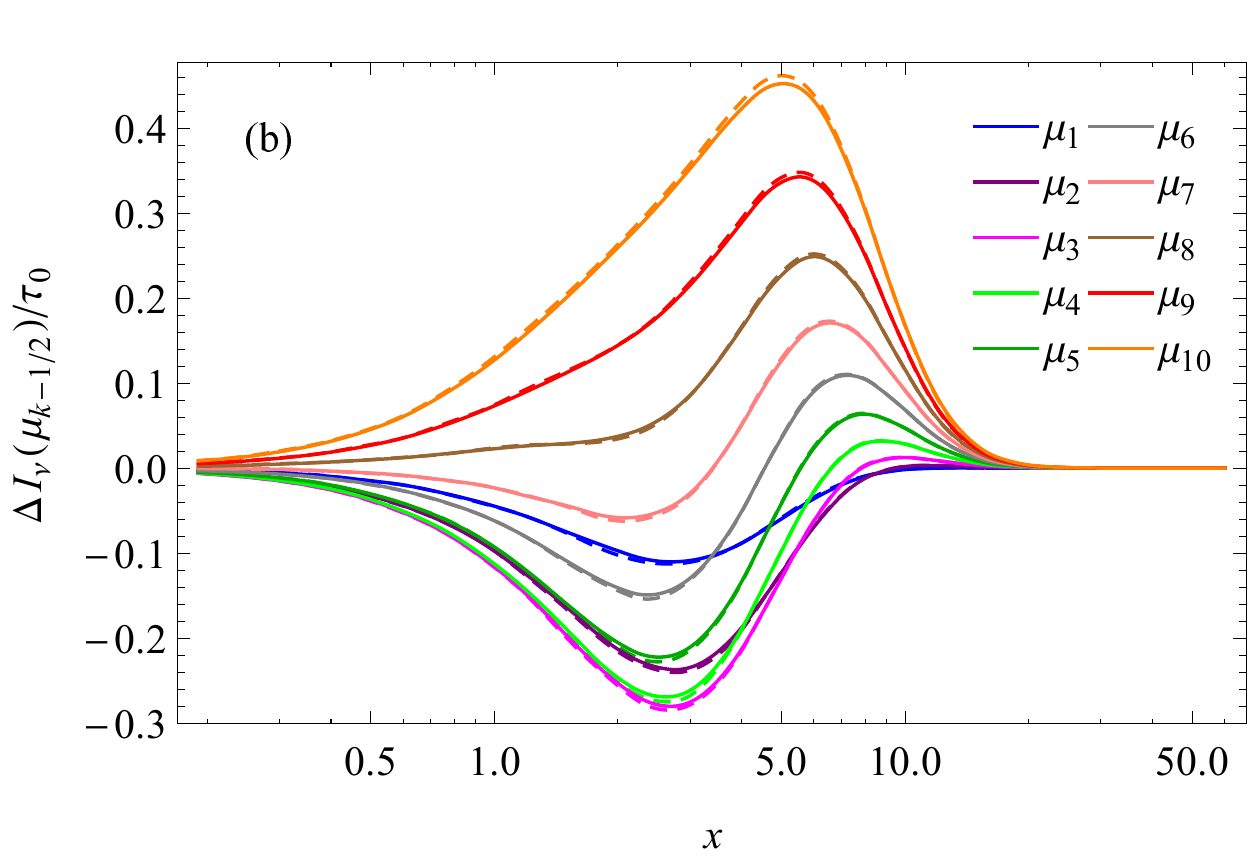}
\hfill
\includegraphics[width=0.45\textwidth]{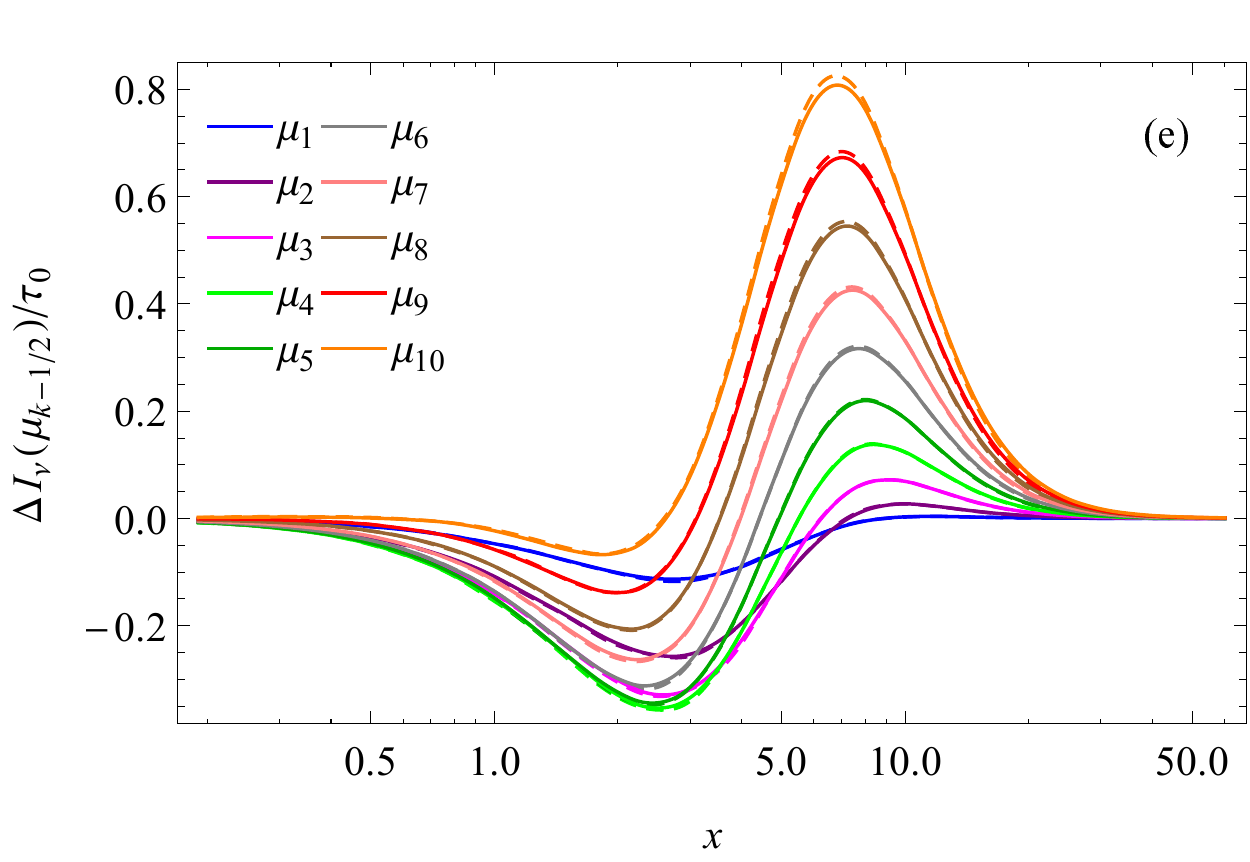}
\vfill
\includegraphics[width=0.45\textwidth]{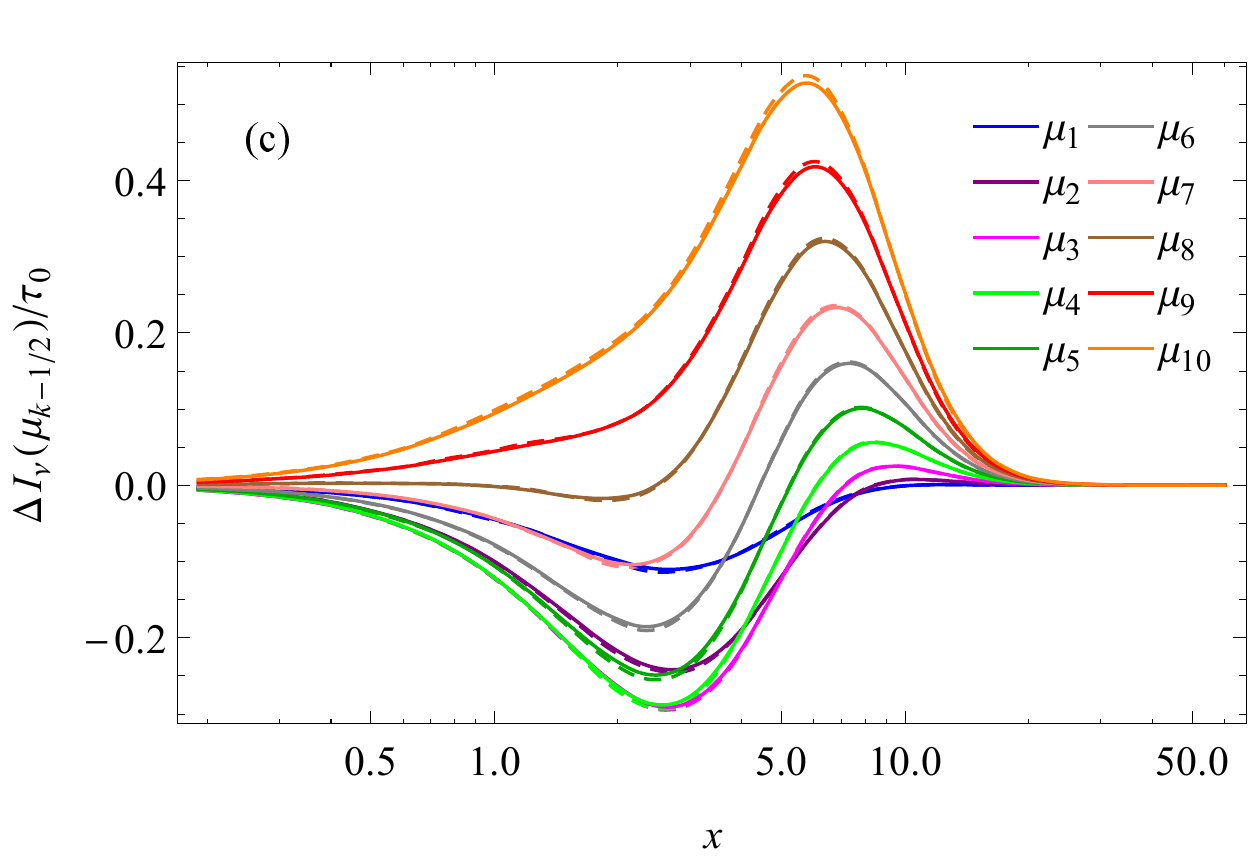}
\hfill
\includegraphics[width=0.45\textwidth]{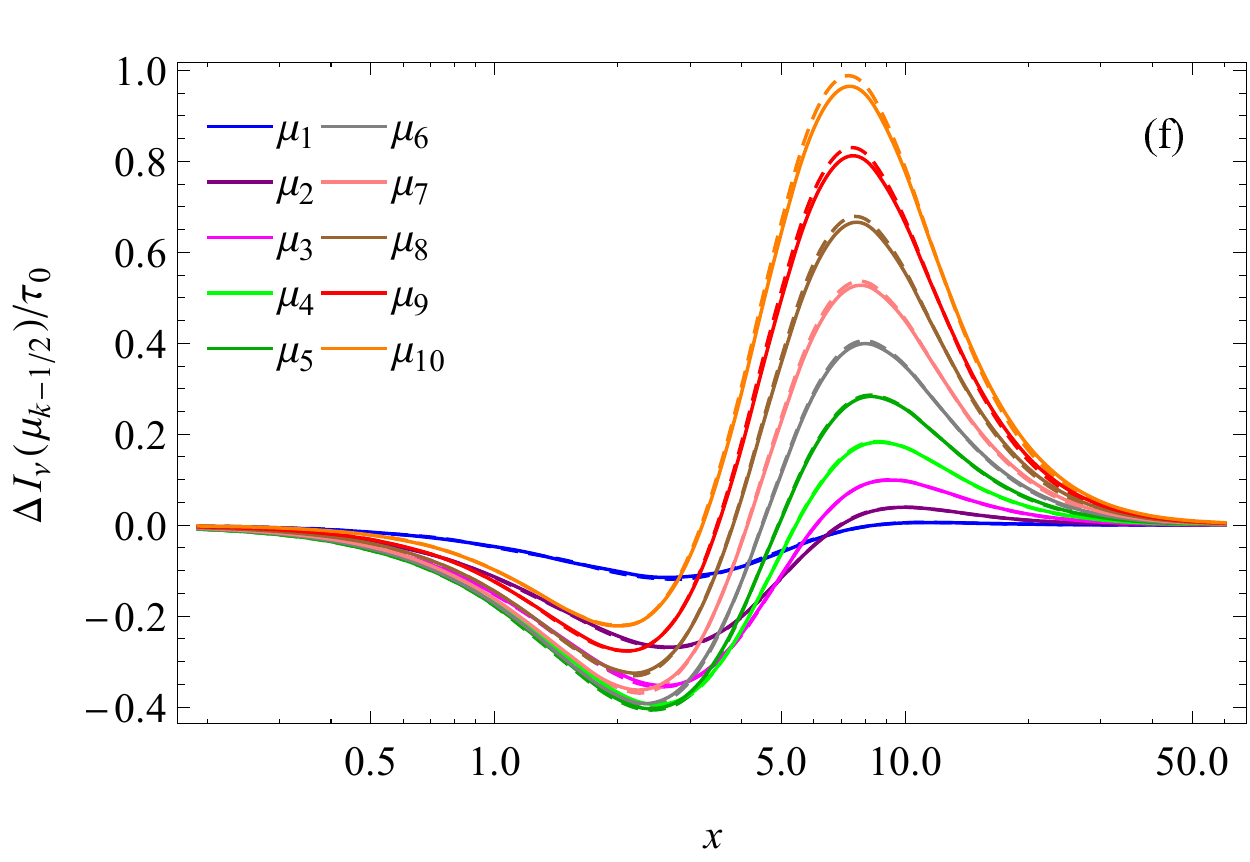}
\end{center}
\caption{The  distortion of spectral intensity of the CMB calculated
for different directions of the photons, outgoing from the cloud, and for the set of values
of temperature $\Theta$: (a) 0.01, (b) 0.02, (c) 0.03, (d) 0.05, (e) 0.07, (f) 0.1.
The optical thickness $\tau_0=0.1$ (solid lines) and $\tau_0=0.05$ (dashed lines), the same colours correspond to the same $\mu$.
}
\label{fig:szangle}
\end{figure*}

\begin{figure*}
\begin{center}
\includegraphics[width=0.45\textwidth]{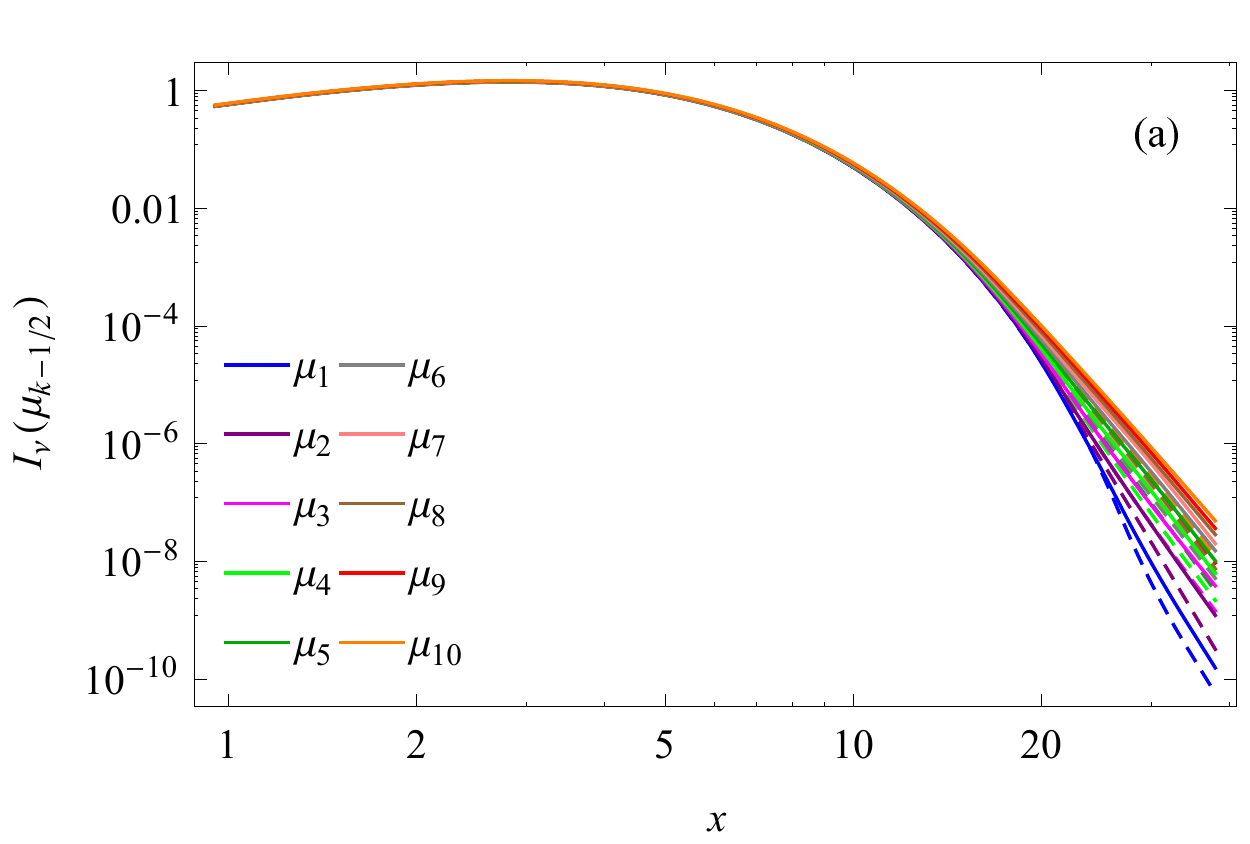}
\hfill
\includegraphics[width=0.45\textwidth]{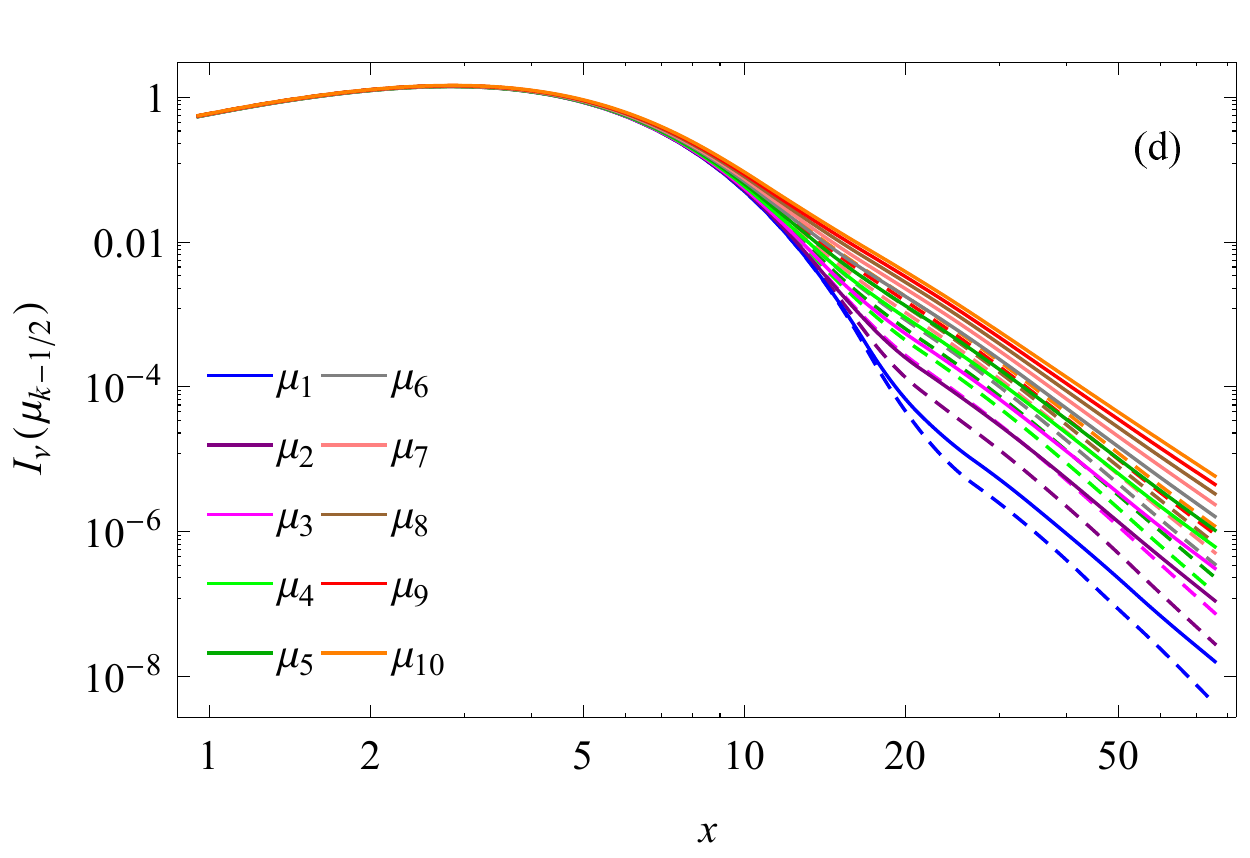}
\vfill
\includegraphics[width=0.45\textwidth]{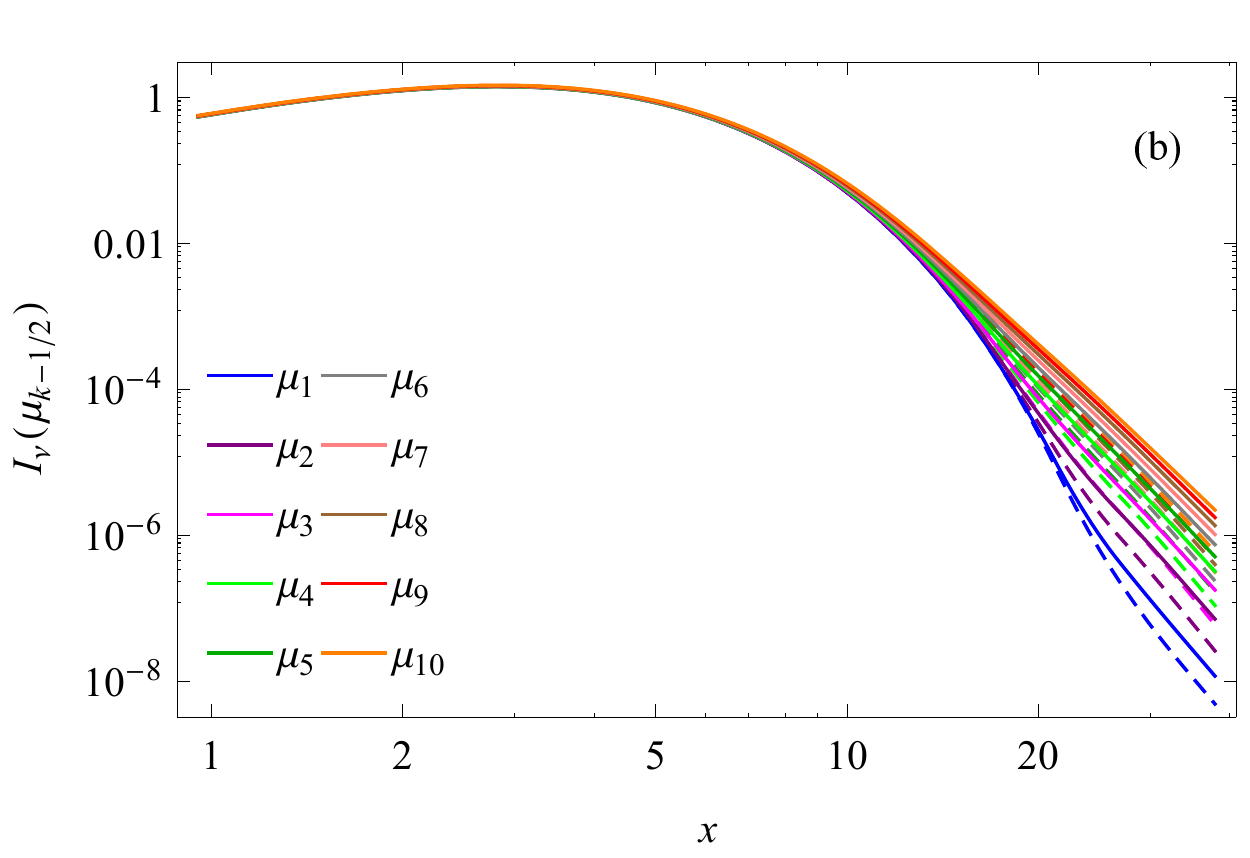}
\hfill
\includegraphics[width=0.45\textwidth]{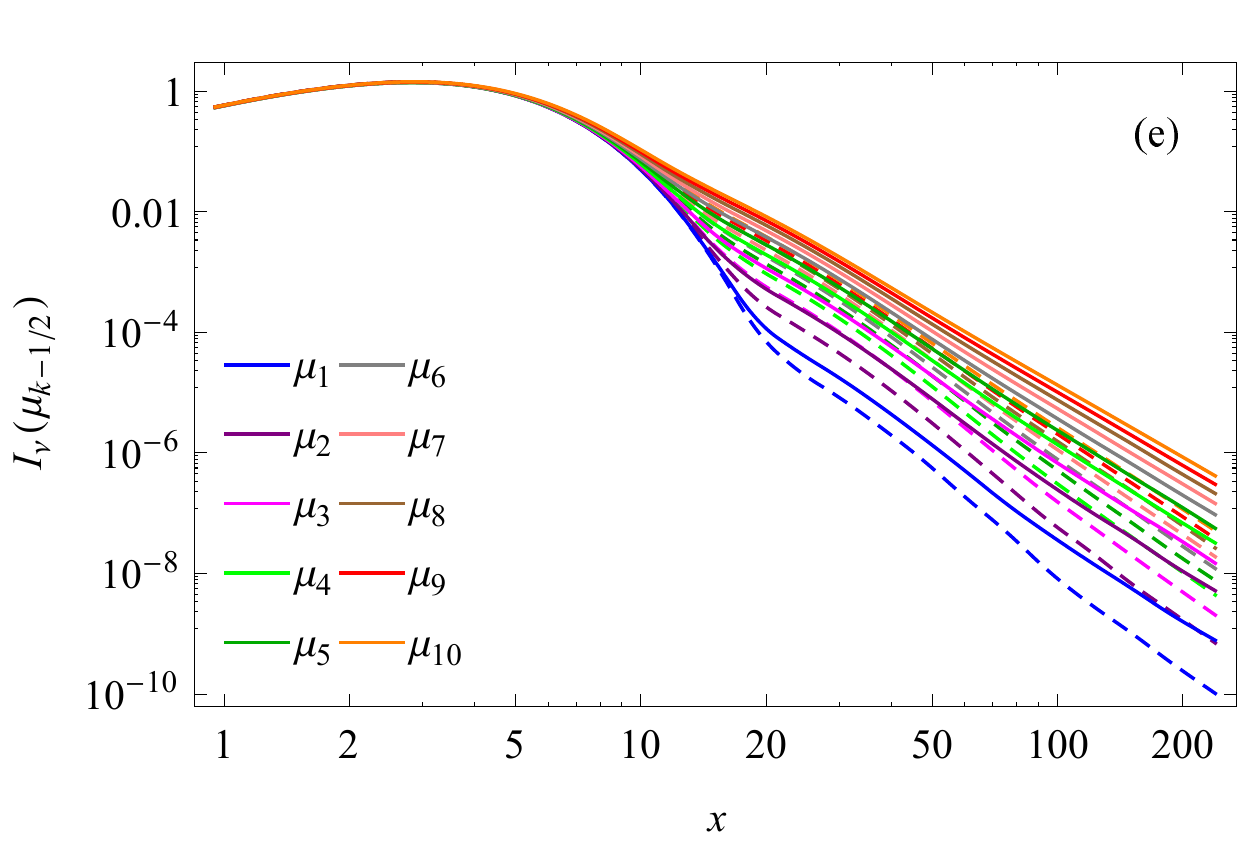}
\vfill
\includegraphics[width=0.45\textwidth]{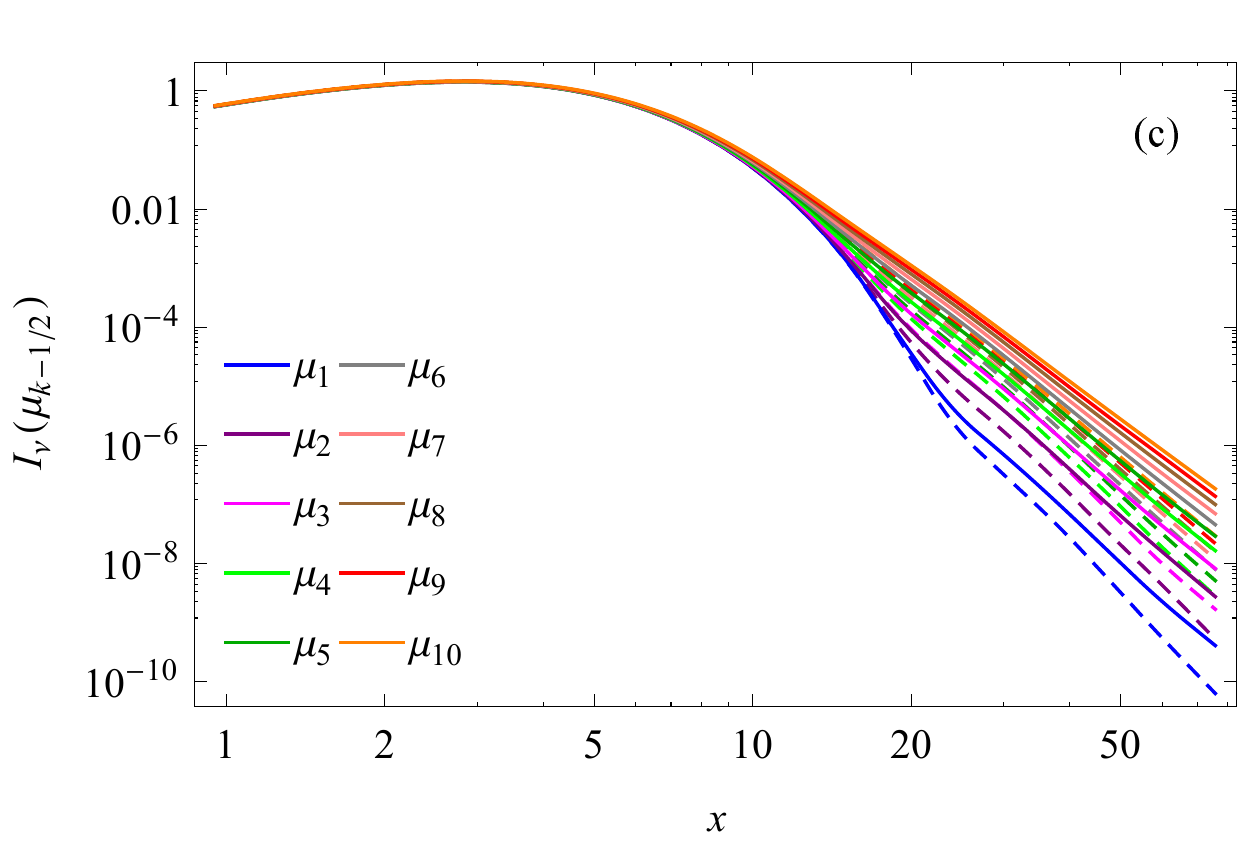}
\hfill
\includegraphics[width=0.45\textwidth]{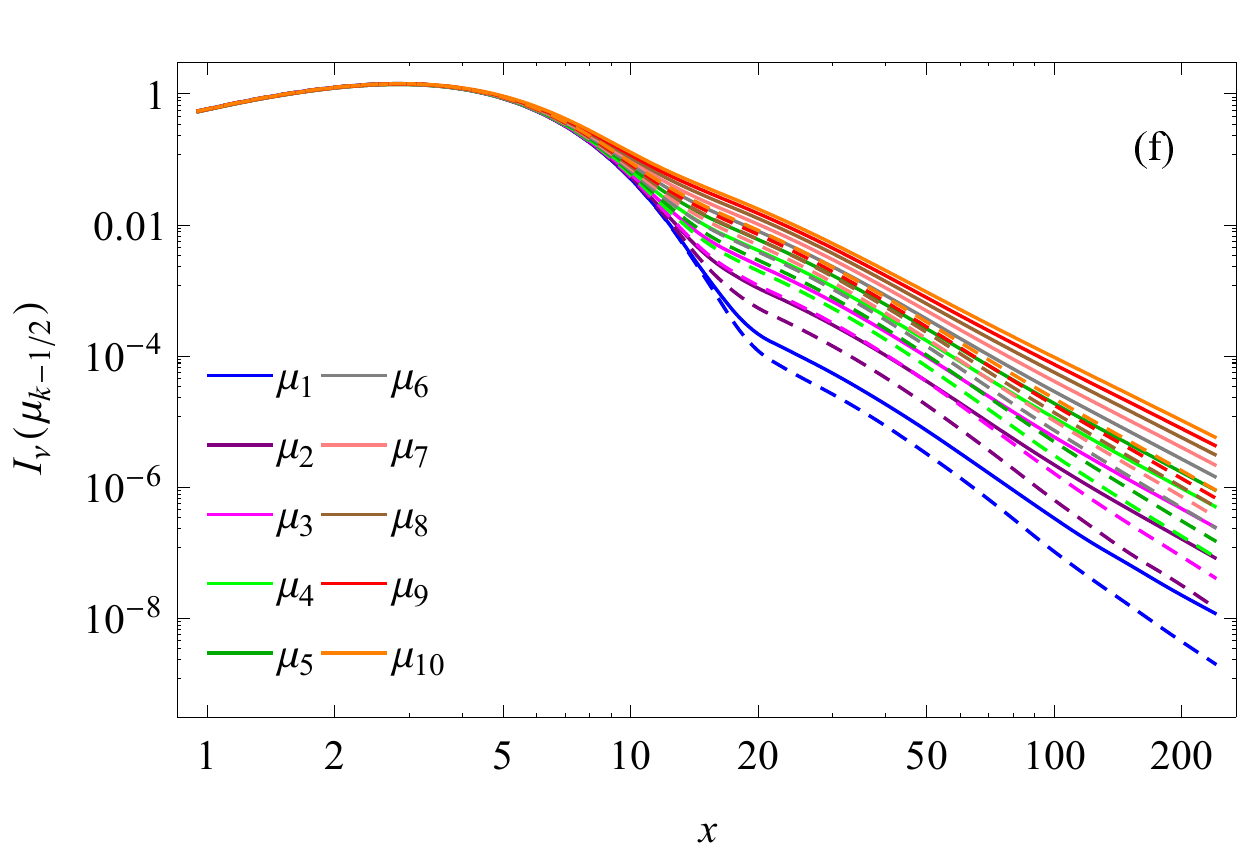}
\end{center}
\caption{Spectral intensity of the CMB, outgoing from the cloud in different directions, calculated
for the set of values of temperature $\Theta$: (a) 0.01, (b) 0.02, (c) 0.03, (d) 0.05, (e) 0.07, (f) 0.1.
The optical thickness $\tau_0=0.1$ (solid lines) and $\tau_0=0.05$ (dashed lines), the same colours correspond to the same $\mu$.
}
\label{fig:szanglesp}
\end{figure*}

Fig. \ref{fig:szangle}  shows the distortions $\Delta I_\nu(\mu)/\tau_0$ (with $\Delta I_\nu(\mu)=I_\nu(\mu)-B_\nu$)
of the spectral intensity
 of the radiation outgoing from the
cloud surface in different directions.  The corresponding spectra are plotted in Fig.~\ref{fig:szanglesp}.
We assume that the intensity is proportional to the calculated quantities $\tilde{I}_k$ taken at the centres of the $\mu$-bins, i.e.  $\tilde{I}_{k-1/2}=\tilde{I}_k(\mu_{k-1/2})\propto I_\nu(\mu_{k-1/2})$, with $\mu_{k-1/2}=(\mu_{k-1}+\mu_k)/2$. The legends
in the panels are displayed in terms of the right boundaries of $\mu$-bins (which are set to be twice wider than in calculations described in \S\ref{ss:res1}).

The linearity of spectral distortion on the optical thickness of the cloud $\tau_0$ holds for each direction (with accuracy demonstrating in Fig. \ref{fig:szangle}).
The spectra at different angles differ markedly
from each other in shape.
The radiation coming out in the directions
corresponding to the values of $\mu$ close to zero
brings the significant contribution to the mean intensity mainly through a slightly distorted (flattened) Planck hump. The spectral distortion
of corresponding spectra is relatively insignificant (Fig.~\ref{fig:szangle}), the intensity of the Comptonization tails has 
less magnitude (Fig.~\ref{fig:szanglesp}) compared with the intensity in other directions at the fixed frequency.
Thus, as one might intuitively expect, at the small values of $\mu$ the radiation has the least distorted spectrum.

The radiation coming out in the directions close to the direction of the outer normal to the cloud boundary ($\mu$ is close to 1)
is characterized by a more pronounced manifestation of the Comptonization effect.
The result of a significant outflow of photons into the Wien region of the spectrum is visible, a power-law tail begins at lower
frequencies (the `bend' disappears) and has a relatively high intensity. Such spectra are formed by photons, which, on average, experience a larger number of scattering events than the average number over the angle-integrated spectrum.

\begin{figure}
\begin{center}
\includegraphics[width=0.45\textwidth]{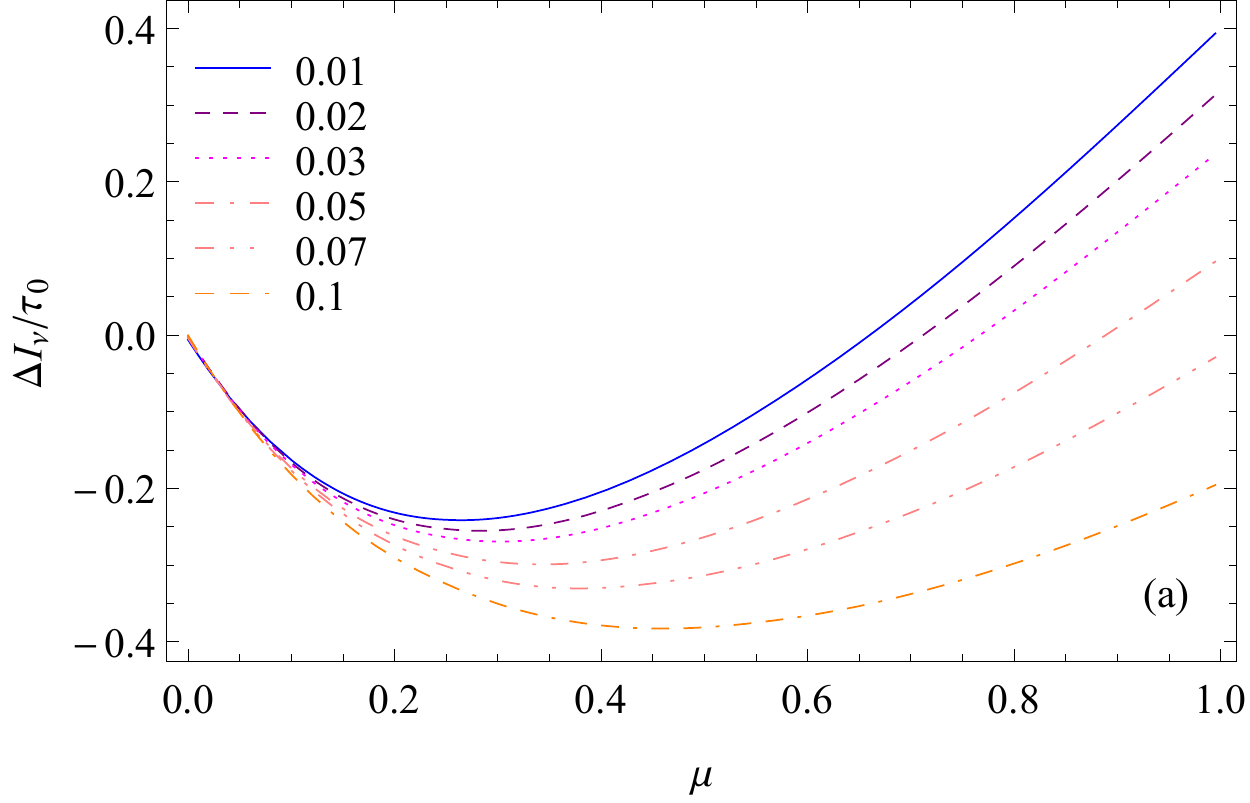}
\vspace{2ex}
\\
\includegraphics[width=0.45\textwidth]{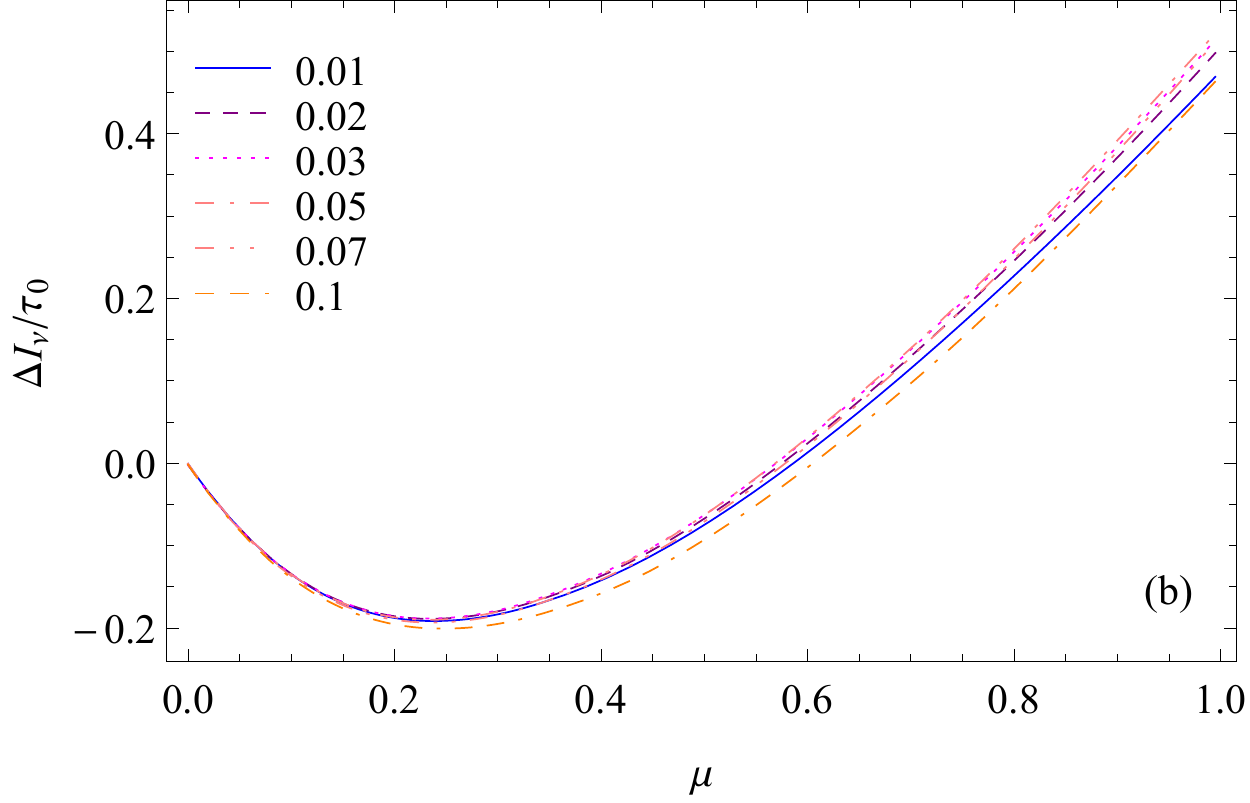}
\vspace{2ex}
\\
\includegraphics[width=0.45\textwidth]{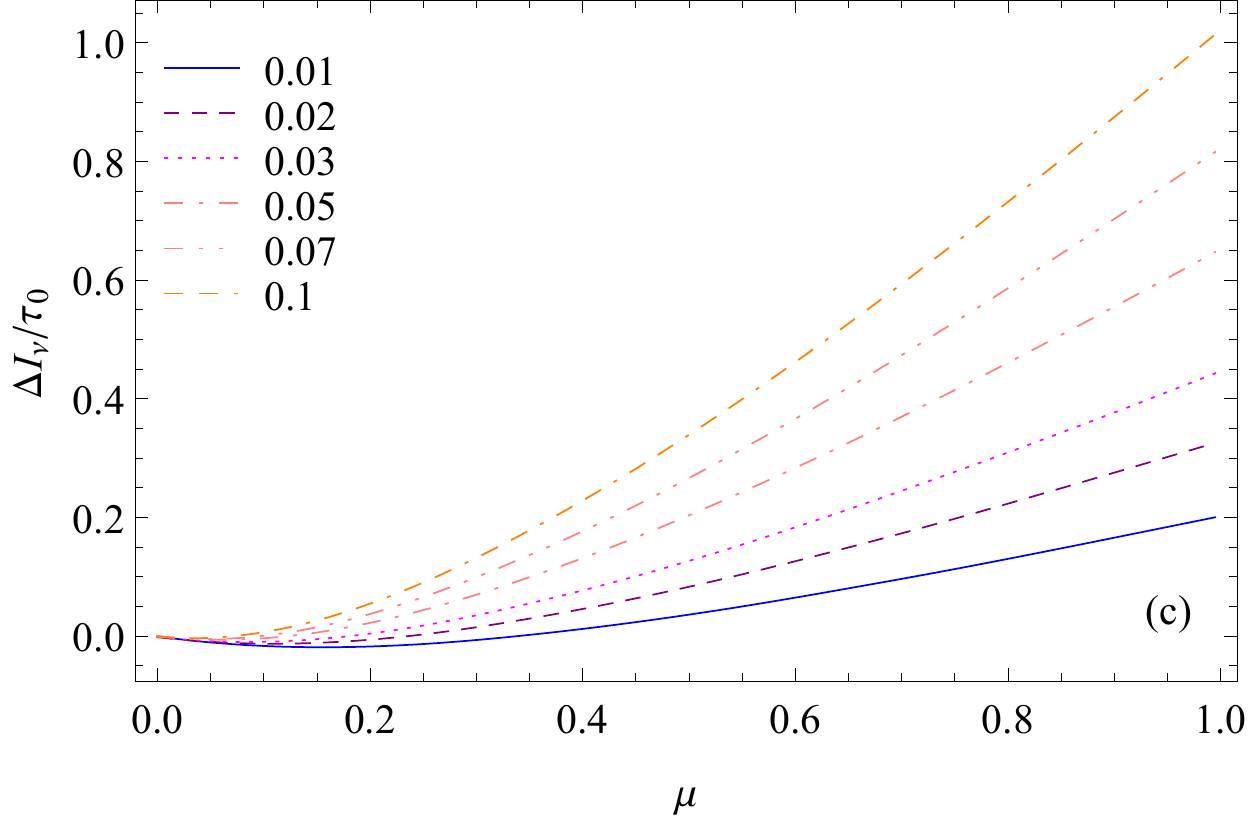}
\end{center}
\caption{Angular distribution of the spectral distortion of the CMB radiation, outgoing from the cloud, calculated for the
photon energy $x=2$ (a), $x=4.3$ (b), $x=8$ (c).
The values of the temperature $\Theta$, corresponding to each curve,  are indicated in the legend. }
\label{fig:szangulardep}
\end{figure}

To describe the results additionally, we have plotted in Fig.~\ref{fig:szangulardep} the angular dependence of the
spectral distortion of the CMB radiation, outgoing from the cloud,
for the different electron temperature and
 for three values of the dimensionless photon energy, $x=2$, $x=4.3$, and $x=8$
(these particular distributions correspond to calculations for $\tau_0=0.1$).
The energies $x=2$ and $x=8$ are not so far from the characteristic energies of the
minimum and maximum of the angle-averaged effect, respectively.
In the vicinity of $x=4.3$, where the angular distributions have the similar form independently from the temperature,
there is a tendency for the curves, corresponding to each $\Theta$, to inverse their  order (cf. Fig.~\ref{fig:szangulardep}a and
Fig.~\ref{fig:szangulardep}c).

\subsection{Applying to observations fitting}
\label{ss:resfit}

The number of CMB experiments (future, under the development or already active) will result in a significant progress in the sensitivity of
observations. At present there are several subarcminute instruments: ALMA and ACA arrays (84-116 GHz),
the bolometer camera MUSTANG-2 on the Green Bank Telescope (90 GHz), the camera NIKA2 on the IRAM telescope, working at frequencies
150 and 260 GHz.  Although the important advances in the  detection of the SZ effect have been achieved with these tools, as well as
in the frames of a number of other experiments (Planck satellite, ACT and SPT telescopes, APEX-SZ, CARMA (SZA), etc.),
there is still ample scope for developing the existing observational base.
The future instrumentation --- AtLAST and CSST telescopes,
CMB-S4 project, Simons Observatory and many others ---
implies increasing available frequency resolution in the wide
band from several tens to several hundreds GHz. The reference information on the projects and
a detailed description can be found, for instance, in a recent review by \cite{2019SSRv..215...17M}.

The wider FoV and higher
spatial resolution will lead to the refinement of description of intergalactic
medium state.
The standard description of tSZ signal is
related with the angle-averaged solutions for the distortion $\Delta J_\nu$ from \S\ref{ss:res1}.
The spatially resolved observations in accordance with our computations are expected to be characterised by the normalized
intensity spectral distortion (caused by tSZ effect) whose
shape depends on the specific direction of line of sight.

The way of the fitting of observational data using the code described in our paper
is determined by the specific aims and characteristic time necessary to achieve them.
The spectra described above (and the corresponding distortion functions) are calculated typically
for the time from a day to several days, depending on the values of the parameters and the task.
The calculations are carried out on the  based on Intel Core i9-9900KF machine,
simultaneously for different sets of the problem parameters (for each set the program is executed by a single thread,
the typical time of computations is specified above under the condition when there is no more than one run per physical core.)

Thus, in order to fit the observations it seems reasonable to  calculate preliminary an extensive set
of spectra for different values of the parameters that are varied with the definite steps,
and further to select the theoretical spectrum which fits the observed one best of all.
If necessary, it is possible to perform additional simulations and make a more
accurate fit around the found first approximation, selecting parameters using a smaller step between its values.
The general regularities inherent in the fitting procedure are well traced from the particular results described above.

\section{Conclusion}
\label{s:conc}
In the  present work we demonstrate the possibilities of our Monte Carlo numerical code for  modelling  the radiation transfer in the hot relativistic plasma with the examples of solving the number of
benchmark problems. We use them to investigate the properties of the thermal Comptonization of the CMB on the electrons of the intergalactic medium of galaxy clusters.

The tSZ effect has been investigated by us in the wide temperature range from $0.01mc^2$ to $0.1mc^2$ and for the different values of the Thomson optical thickness of the cloud. The CMB spectra, including their power-law tails, are calculated with insignificant errors up to energies $x\gtrsim 80$ for most computations.

Several test problems solved here confirm the results of the previous researches, showing  at the same time the limits of their  applicability.
The test problem to find spectra of the once-scattered photons has been solved and demonstrated  the physical adequacy of our numerical realisation of the relativistic Compton scattering kernel.

The computed line profiles of the initially monochromatic soft photons with energy  $h\nu_0\ll k_{\rm B}T_{\rm e}$  show that the spectrum slope in the vicinity of the source differs from its value at $\nu\gg\nu_0$, where the spectral distribution is a power-law.
The fitting of the line profiles by the broken power-law Green function (\ref{e:GF}) leads to the conclusion that the spectral indexes $\alpha_-$ and $\alpha_+$ at the different sides of the line frequency  are related as $\alpha_-=\alpha_++\varsigma$, where $\varsigma\sim 3$ even in the case of the optically thin medium.
The spectral index $\alpha$, obtained by fitting power-law intervals of the spectra of monochromatic line,
is very close to the one found from \eq{e:disp}.
The power-law tails of the numerically calculated Comptonized CMB spectra have the spectral indexes satisfying the same equation.

Considering a uniform spherical plasma cloud we have obtained the angular dependence of the outgoing distorted CMB radiation.
The spectral shape of the intensity distortion, caused by tSZ effect, is shown to be not universal
for different directions of the photons escaping from the cloud.
This result, being a consequence of the treatment of the radiation transfer in a spatially bounded medium,
complements the previous approaches and can be considered in the context of the future studies of the physics of galaxy clusters.
Evidently, our method allows us to consider a more involved  problem, for example, to find a solution in the case of a non-zero
peculiar velocity of a galaxy cluster or when its geometry and/or density distribution has a complicated character. The
code can be considered as a useful tool to study a distribution of the intergalactic gas in galaxy clusters,
and also for testing the cosmological models.

\section*{Acknowledgements}
We appreciate Prof. L.\,G.\,Titarchuk for the discussions.
Furthermore, we thank Prof. Yu.\,E.\,Lyubarskij for the important remarks.
MG acknowledges Prof. K.\,A.\,Postnov for the attention to the text and useful comments.
The work of MG on  calculation of Comptonization of the CMB is supported by the grant 17-15-506-1 of the
Foundation for the advancement of theoretical physics and mathematics `BASIS'.
The work of MG on calculation of Comptonization of monochromatic lines and single-scattering profiles
was funded by RFBR according to the research project 18-32-00890.
The research of analytical solutions for the spectral index
was supported by the Program of development of M.\,V.\,Lomonosov
Moscow State University (Leading Scientific School `Physics of stars, relativistic objects and galaxies').
GVL acknowledges the support by the Program of development of M.\,V.\,Lomonosov Moscow State University.

\section*{Data availability}
The calculations described in this paper were performed using the private codes developed by the corresponding author.
The data presented in the figures as well as additional information are available on reasonable request.

\bibliographystyle{mnras}
\bibliography{cmb}

\label{lastpage}
\end{document}